\documentclass[aip,cha,reprint,amsmath,amssymb,floatfix]{revtex4-1}

 \pdfoutput=1 

\usepackage[version=4]{mhchem}
\usepackage{units}
\usepackage{xcolor}
\usepackage{graphicx}
\usepackage{upgreek}
\usepackage{bbold}
\usepackage{booktabs}
\usepackage{tablefootnote}
\newcommand{\ra}[1]{\renewcommand{\arraystretch}{#1}}
\usepackage{braket}
\usepackage[utf8]{inputenc}
\usepackage[T1]{fontenc}

\usepackage[hidelinks]{hyperref}
\makeatletter
\let\oldtheequation\theequation
\renewcommand\tagform@[1]{\maketag@@@{\ignorespaces#1\unskip\@@italiccorr}}
\renewcommand\theequation{(\oldtheequation)}
\makeatother

\newcommand{\deriv}{\mathrm{d}}  

\begin{document}


\title{Designing Broadband Pulsed Dynamic Nuclear Polarization Sequences in Static Solids} 




\author{Nino Wili}
\email{nino.wili@alumni.ethz.ch}
\affiliation{Department of Chemistry and Applied Biosciences, Laboratory of Physical Chemistry, ETH Zurich,
	Vladimir-Prelog-Weg 2, 8093 Zurich, Switzerland}
\author{Anders Bodholt Nielsen}
\affiliation{Interdisciplinary Nanoscience Center (iNANO) and Department of Chemistry, Aarhus University, Gustav Wieds Vej 14, DK-8000 Aarhus C, Denmark}

\author{Laura Alicia Völker}
\author{Lukas Schreder}
\affiliation{Department of Chemistry and Applied Biosciences, Laboratory of Physical Chemistry, ETH Zurich,
	Vladimir-Prelog-Weg 2, 8093 Zurich, Switzerland}

\author{Niels Chr. Nielsen}
\affiliation{Interdisciplinary Nanoscience Center (iNANO) and Department of Chemistry, Aarhus University, Gustav Wieds Vej 14, DK-8000 Aarhus C, Denmark}

\author{Gunnar Jeschke}
\affiliation{Department of Chemistry and Applied Biosciences, Laboratory of Physical Chemistry, ETH Zurich,
	Vladimir-Prelog-Weg 2, 8093 Zurich, Switzerland}

\author{Kong Ooi Tan}
\email{kong-ooi.tan@ens.psl.eu}
\affiliation{Laboratoire des Biomolécules, LBM,  Département de Chimie, École Normale Supérieure, PSL University, Sorbonne Université, CNRS, 75005 Paris, France}


\date{\today}

\begin{abstract}
Dynamic nuclear polarization (DNP) is an NMR hyperpolarization technique that mediates polarization transfer from highly polarized unpaired electrons to NMR-active nuclei via microwave (mw) irradiation. The ability to generate arbitrarily shaped mw pulses using arbitrary waveform generators opens up the opportunity to remarkably improve the robustness 
and versatility of DNP, in many ways resembling the 
early stages of pulsed NMR.
We present here novel design principles based on 
single-spin vector effective Hamiltonian theory to develop
new broadband DNP pulse sequences, namely an adiabatic XiX-DNP experiment and a 
broadband amplitude modulated signal enhanced (BASE) experiment. We demonstrate that the adiabatic BASE 
pulse sequence may achieve a DNP $^{1}$H enhancement factor of $\sim$ 360, a record that outperforms all previously known pulsed DNP sequences at $\sim$ 0.35 T and 80 K in static solids. 
The bandwidth of the BASE-DNP experiments is about 3 times the $^{1}$H Larmor frequency ($\sim$50 MHz).
\end{abstract}


\maketitle 



%
%

%

\section{Introduction}

Dynamic nuclear polarization (DNP) is a powerful tool to increase the sensitivity of nuclear magnetic resonance (NMR) by transferring the much higher polarization 
of electron spins to nuclear spins with a theoretical maximum 
enhancement factor $\varepsilon\sim658$ 
\cite{Abragam1978,LillyThankamony2017} for $^{1}$H. The hyperpolarization method
allows one to study systems that suffer from poor NMR sensitivity with reduced measurement time or cost. For instance, a DNP experiment with $\varepsilon\sim100$ performed in one hour would have 
taken $\sim 1$ year without mw irradiation for the same-quality 
spectrum. This opens up the possibility to extract important structural information from small molecules, 
biological samples, or inorganic materials that are otherwise inaccessible due to poor NMR sensitivity.

Since the discovery of DNP in the fifties \cite{Overhauser1953}
tremendous progress has been made, and there are two main DNP methods: dissolution DNP \cite{Ardenkjaer-Larsen2003} and 
\textit{in situ} solid-state DNP NMR potentially combined with magic-angle spinning (MAS) \cite{Becerra1993}. 
The former category typically polarizes a static sample at low temperatures (< 2 K) and moderate magnetic fields (3.4-10.1 T)\cite{ArdenkjrLarsen2018}, where the electron polarization approaches unity.  
Following that, the sample undergoes a dissolution process prior to be transported to a high-resolution NMR magnet or MRI system for detection in solution state. The \textit{in situ} static sample or MAS DNP NMR approach performs the hyperpolarization and NMR detection process on solid samples in the same high-resolution magnet typically at temperatures below 100 K. To facilitate efficient DNP processes at these relatively higher temperatures and fields, it is necessary to use high-power mw sources known as gyrotrons. Although these DNP methods have successfully enabled many biological molecules and materials to be studied with superior sensitivity \cite{LillyThankamony2017,Jannin2019a}, it is noted that the DNP efficiency deteriorates at higher magnetic fields \cite{Tan2019a}. Despite new classes of biradicals that have been reported to circumvent this issue\cite{Berruyer2020,Cai2021}, the DNP performance is sample-dependent and varies with rotor sizes (or mw penetration). Hence, it would be highly desirable to develop a general DNP method with a consistent performance across different magnetic fields, paramagnetic polarizing agents, and other experimental conditions. One of the major reasons that hinders the development towards this goal is that most contemporary DNP approaches use continuous-wave (CW) mw irradiation, where the amplitude, phase, and frequency cannot be easily controlled. As a consequence only four main CW-DNP mechanisms have been discovered so far, namely: Overhauser effect (OE) \cite{Overhauser1953}, the solid effect (SE) \cite{Abragam1958,C.D.Jeffries1957}, the cross effect (CE) \cite{Hwang1967,Kessenikh1963}, and thermal mixing (TM) \cite{Provotorov1962,Borghini1968}. In comparison, hundreds of NMR pulse sequences have been invented to date, for purposes ranging from polarization transfer, distance measurement, to determination of dynamics and chemical environments, etc. An important aspect in choosing the right pulse sequence is the bandwidth, an issue relatively less discussed in DNP. The EPR line widths of many radicals are often broad, especially at high fields due to large g-anisotropy. The linewidth of a typical nitroxide radical is $\sim$ \unit[1]{GHz} at \unit[9.4]{T},\cite{Soetbeer2018} which is orders of magnitude higher than the electron Rabi fields conferred by the currently available mw power. The aim of this study is to demonstrate - albeit at lower field - ways to design broadband and efficient  DNP techniques exploiting shaped pulses generated by arbitrary waveform generators (AWG) at different mw power conditions.

Our approach to the design of broadband pulsed DNP experiments has been inspired by previous pulsed DNP techniques, namely nuclear orientation via spin locking (NOVEL) \cite{Henstra1988,Can2015a,Mathies2016}, 
ramped-amplitude (RA)-NOVEL \cite{Can2017b}, off-resonance NOVEL \cite{Jain2017}, the integrated solid effect (ISE) \cite{Henstra1988a,Can2017}, the adiabatic solid effect (ASE) \cite{Tan2020}, nuclear rotating frame (NRF)-DNP \cite{Wind1988}, 
the dressed spin solid effect (DSSE) \cite{Weis2000}, PulsePol \cite{Schwartz2018a}, and time-optimized pulsed (TOP)-DNP \cite{OoiTan2019}. 
Among these sequences, we would like to emphasize that the TOP-DNP sequence is substantially different from others, i.e., the initial truncation of 
the electron-nuclear dipolar couplings by the nuclear Zeeman term can be reintroduced by mw irradiation to transfer polarization in DNP.
The methods to reintroduce  these couplings are mathematically similar to and inspired by the 
dipolar recoupling techniques in MAS solid-state NMR spectroscopy, where rf irradiation interferes with the rotational averaging of dipolar couplings. Following this, a phase-alternating low-power X-inverse X (XiX)-DNP has been demonstrated recently using similar average Hamiltonian and operator-based Floquet theory design principles \cite{Mathies2022}.  In this work, we further introduce an alternative design strategy based on single-spin vector effective 
Hamiltonian theory \cite{Shankar2017,Nielsen2019}, which incorporates Fourier coefficients (exploited in Floquet theory) into average Hamiltonian theory. This theoretical framework is applicable to \textit{any} arbitrary periodic DNP experiments, and will here be used for designing broadband DNP experiments. We will examine these sequences by  numerical simulations and experiments at \unit[0.35]{T}/ \unit[9.8]{GHz}/ \unit[15]{MHz} on OX063 trityl radicals doped in a glycerol-water mixture at 80 K.

\section{Theory}

We will  describe the overall Hamiltonian 
for an electron-nuclear spin system followed by a series of 
transformations leading to a convergent effective Hamiltonian. The 
Hamiltonian is cast as a Fourier expansion, which allows us to  identify the resonance conditions and determine the effective couplings that govern the DNP polarization transfer. The procedure takes inspiration from previous 
single-vector effective Hamiltonian approaches described recently in 
relation to solid-state NMR dipolar recoupling \cite{Shankar2017} and liquid-state NMR 
isotropic mixing \cite{Nielsen2019}.

While the theory outlined in the following is valid and extendable to 
describe systems with multiple electrons and nuclei,
we will for simplicity stick to a two-spin system comprised of one 
electron spin ($S$) 
and one nuclear spin ($I$). The laboratory-frame Hamiltonian is given by
\begin{align}
	\mathcal{H} =  \omega_S S_z + \mathcal{H}_\text{mw} + 
	\vec{S} \cdot \mathbf{A} \cdot \vec{I}+ \omega_I I_z 
	\quad,
	\label{eq:lab_frame_ham}
\end{align}
with $\omega_S = -\gamma_\text{e} B_0$ and $\omega_I = 
-\gamma_\text{n} B_0$ being angular frequencies for the electron and 
nuclear Zeeman interactions, respectively (for an e-$^{1}$H system 
$\omega_S > 0$ and $\omega_I$<0). $\gamma$, $\mathbf{A}$, $\mathcal{H}_\text{mw}$, and $B_0$ refer to the gyromagnetic ratio, 
the hyperfine coupling tensor, the Hamiltonian of the mw 
irradiation, and the external static magnetic field along the $z$-axis, respectively. Upon transformation to the electron 
rotating frame and employing the high-field 
approximation , the first-order effective Hamiltonian becomes
\begin{equation}
	\bar{\tilde{\mathcal{H}}} =  \underbrace{ \Omega_S S_z+\tilde{\mathcal{H}}_\text{mw} 
	}_{\mathcal{H}_{\text{control}}} + \underbrace{ A_{zz}S_zI_z  + 
		B S_z I_x +\omega_I I_z }_{\mathcal{H}'} \quad,
	\label{eq:rot_frame_ham1}
\end{equation}
where  $\Omega_S = \omega_S -\omega_\text{mw}$ is the mw offset frequency;  $A_{zz}$ and 
$B=\sqrt{A_{zx}^2 + A_{zy}^2}$ are the secular and pseudosecular coupling, respectively. Note that the $B$ term originates purely from the dipolar coupling, which is averaged to zero in solution state. We will restrict our discussions to spin systems in the regime where
$|A_{zz}|,|B|\ll |\omega_I|$, so that the effect of the spin diffusion barrier can be neglected in this context, i.e. the electron-nuclear distance falls in the 
range of \unit[4-10]{\AA} \cite{Wolfe1973,Tan2019,Jain2021,Stern2021}. 

We now move into the interaction frame with the control field 
$\mathcal{H}_\text{control}$, which contains all  explicit 
information (amplitudes, frequencies, and phases) about mw pulses:
\begin{align}
	\tilde{\mathcal{H'}} =& U_{\text{control}}^\dagger \, \mathcal{H'} \, 
	U_{\text{control}} \nonumber\\
	=& \sum_{\chi=x,y,z} R^{(\text{control})}_{\chi z} (t) S_\chi (A_{zz} 
	I_z+B I_x) + \omega_I I_z \quad,
	\label{eq:rot_frame_ham2}
\end{align}
where $U_{\text{control}}(t)=\hat{T}\exp{(-i \int_{0}^{t} 
	\mathcal{H}_\text{control}(\tau)\mathrm{d}\tau)}$ and $\hat{T}$ is the Dyson 
time-ordering operator. The time-dependent rotation matrix $R^{(\text{control})}_{\chi z}$ represents the 
interaction-frame trajectory of the electron spin under the mw pulse sequence. Then,  we can calculate  an effective time-independent Hamiltonian using standard 
average Hamiltonian theory, provided $\mathcal{H}_\text{control}(t)$ 
is periodic over a given period $\tau_\text{m} =2 \pi/\omega_\text{m}$ (where $\omega_\text{m}$ is the 
modulation frequency of the pulse sequence), i.e.,  
$\mathcal{H}_\text{control}(t)$ = 
$\mathcal{H}_\text{control}(t+\tau_{m})$. We note that 
a periodic Hamiltonian does not necessarily need to have an identity propagator over one cycle.

Although the control propagator  ${U}_\text{control}$, which is equivalent to $R_{\chi z}^\text{(control)}$, may not necessarily be cyclic in the normal mw rotating frame, it is always possible to find a frame in which the new transformed control propagator becomes cyclic. Having a cyclic propagator is particularly useful in both characterizing and gaining useful insight into the pulse sequence. It is also a prerequisite for the application of average Hamiltonian theory to the interaction frame Hamiltonian. Hence, we choose a frame with its $z$-axis aligned with the effective field ($\omega_{\text{eff}}$)
\cite{Shankar2017,Nielsen2019}, whose magnitude and direction can be determined using quaternion algebra \cite{Blumich1985,Counsell1985,Tan2015a}. We indicate this frame in the following with a tilde on the $\tilde{S}$-spin operators. This leads to
\begin{equation}
	\tilde{\mathcal{H'}} = 
	\sum_{\chi=x,y,z} R^{(\text{eff})}_{\chi z} (t) \tilde{S}_\chi (A_{zz} 
	I_z+B I_x) - \omega_{\text{eff}}^{(S)} \tilde{S}_{z} + \omega_I I_z
	\label{eq:rot_frame_ham3}
\end{equation}
with 
\begin{align}
	R^{(\text{eff})}_{\chi z}(t)   =& [R_z 
	(-\omega_\text{eff}^{(S)} t) \cdot R^{(\text{flip})}(\beta) \cdot 
	R^{(\text{control})}]_{\chi z}(t)\nonumber \\
	=&\sum\limits_{k=-\infty}^{\infty}a_{\chi z}^{(k)}e^{ik\omega_\text{m}t} 
	\label{eq:rot_frame_ham3b}
\end{align}
representing a three-step transformation comprised of (1) go into the 
control frame; (2)  flip the coordinate system by $\beta$ so that it is aligned 
with $\omega_{\text{eff}}^{(S)}\tilde{S}_z$; (3) rotate the frame by an angle 
$\omega_{\text{eff}}^{(S)}t$ around the new $z$ axis. \cite{Shankar2017,Nielsen2019}.  Moreover, Eq. \ref{eq:rot_frame_ham3b} is cyclic, i.e., $[R^{(\text{eff})}_{\chi z} 
(t+\tau_\text{m})]_{\chi z} =	R^{(\text{eff})}_{\chi z}(t) $, and the elements of the overall rotation matrix may be expressed in terms of a Fourier series as given in the right-hand side of the equation. The term
$-\omega_\text{eff}^{(S)}\tilde{S}_z$ incorporates the Coriolis term originating from step (3) above.

We transform the nuclear part ($I$) of the Hamiltonian into an interaction frame with closest match between the mw modulation frequency $\omega_\text{m}$ and $\omega_I$ to obtain the $I$-spin effective field
\begin{equation}
	\label{eq:nuclear effective field}  
	\omega_\text{eff}^{(I)} = \omega_I -k_I\omega_\text{m} \text{ with }
	k_I = \mathrm{round}\left(\frac{\omega_I}{\omega_\text{m}}\right) 
	,
\end{equation}
which is then used to form the propagator  $U_{\text{eff}}=\exp{\left(-ik_{I} \omega_{\text{m}}t I_{z}\right)}$ that transforms $\tilde{\mathcal{H'}}$ (Eq. \ref{eq:rot_frame_ham3}) into
\begin{align}
	\tilde{\tilde{\mathcal{H}}}' &=  U_{\text{eff}}^\dagger \, \tilde{\mathcal{H'}} \, 
	U_{\text{eff}} - k_I\omega_\text{m}I_z \nonumber\\
	&= \sum\limits_{\chi=x,y,z}\sum\limits_{k=-\infty}^{\infty}a_{\chi z}^{(k)}e^{ik\omega_\text{m}t} \tilde{S}_\chi \times \nonumber \\ 
	&\left(A_{zz} I_z+\frac{B}{2}\left(e^{ik_I\omega_\text{m}t} I^+ + 
	e^{-ik_I\omega_\text{m}t} I^-\right)\right) \nonumber \\
	&-\omega_\text{eff}^{(S)} \tilde{S}_z   + \omega_\text{eff}^{(I)} I_z 
	\label{eq:rot_frame_ham4}
	\quad .
\end{align}
Next, we  apply first-order average Hamiltonian theory (AHT) and recognize that index $k$ in Eq. \ref{eq:rot_frame_ham4} has to be either $=\pm k_I$ or $=0$. All other terms vanish upon integration over one modulation period, leading to 
\begin{widetext}
\begin{align}
	\bar{\tilde{\tilde{\mathcal{H}}}}^{(1)} &= 
	A_{zz}\sum\limits_{\chi=x,y,z}a_{\chi z}^{(0)} \tilde{S}_{\chi} I_z + 
	\frac{B}{2} \tilde{S}_z \left( a_{zz}^{(-k_I)} I^+ + a_{zz}^{(k_I)} I^-\right) \nonumber \\
	&+ \frac{B}{4}\left( a_{+z}^{(-k_I)}\tilde{S}^+I^+ + 
	a_{-z}^{(-k_I)}\tilde{S}^-I^+ + a_{+z}^{(k_I)}\tilde{S}^+I^- + a_{-z}^{(k_I)}\tilde{S}^-I^- \right)  - \omega_\text{eff}^{(S)}\tilde{S}_z +\omega_\text{eff}^{(I)} I_z \quad ,
	\label{eq:full first order}
\end{align}
\end{widetext}
where we define $a_{xz}^{(q)}\tilde{S}_{x}+a_{yz}^{(q)}\tilde{S}_{y} = 
\frac{1}{2} ( a_{+z}^{(q)}\tilde{S}^{+}+a_{-z}^{(q)}\tilde{S}^{-} )$ and 
$a_{\pm z}^{(q)}=a_{xz}^{(q)} \mp ia_{yz}^{(q)}$. Since the effective Hamiltonian must be Hermitian, it is enforced that
$a_{-z}^{(q)}=\left(a_{+z}^{(-q)}\right)^*$.

\subsection{Identifying resonance conditions and scaling factors}
We note that the effective Hamiltonian $\bar{\tilde{\tilde{\mathcal{H}}}}^{(1)}$ (Eq. \ref{eq:full first order}) was derived without explicitly describing the details of the pulse sequence, i.e., it has a general form and is applicable to \emph{all} periodic EPR or DNP sequences acting on a two-spin electron-nucleus system in the regime where $|A_{zz}|,|B|\ll|\omega_I|$. This is possible because the details of these pulses sequences are implicitly encoded in the scaling factors $a_{\chi z}$ of the corresponding resonance conditions, which will be discussed in the following.

For polarization-transfer experiments, it is necessary to retain only either the zero- or double-quantum (ZQ or DQ) operators and suppress all other non-commuting operators. This is achieved by matching the effective fields  ($-\omega_\text{eff}^{(S)}\tilde{S}_z,\omega_\text{eff}^{(I)} I_z$) in Eq. \ref{eq:full first order}, i.e., by finding the DNP matching conditions. One can identify a resonance condition $\omega_\text{eff}^{(S)}=-\omega_\text{eff}^{(I)}$, where only the ZQ operators ($\tilde{S}^\pm I^\mp $) survive because the DQ terms ($\tilde{S}^\pm I^\pm $) are truncated by the larger non-commuting $I_{z}+\tilde{S}_{z}$ term. Note that a converse case is found in the other resonance condition, $\omega_\text{eff}^{(S)}=\omega_\text{eff}^{(I)}$, where only the DQ terms remain due to a similar truncation in the ZQ subspace.  All other terms can be neglected as long as they are much smaller than the effective fields. This is a good approximation for weakly coupled protons involved in DNP, and will be checked by numerical simulations below. While the $A_{zz} \tilde{S}_z I_z$ term commutes with the effective fields,  it  shifts both energy levels in the same direction within the respective subspace (ZQ or DQ). Thus, the energy difference and resonance conditions remain unchanged.
By neglecting the the terms discussed above, we obtain
\begin{align}
	\bar{\tilde{\tilde{\mathcal{H}}}}^{(1)} =& 
	\frac{B}{4}\left(    a_{-z}^{(\mp k_I)}\tilde{S}^-I^\pm   +  
	a_{+z}^{(\pm k_I)}\tilde{S}^+I^\mp \right) \nonumber\\
	    -&\omega_\text{eff}^{(S)}\tilde{S}_z
	+\omega_\text{eff}^{(I)} I_z \quad \text{for} \quad 
	\omega_\text{eff}^{(S)}\approx\mp\omega_\text{eff}^{(I)} \quad ,
	\label{eq:truncated first order ZQ}
\end{align}
and the  $\rho_0\rightarrow I_{z}$ transfer mediated by the effective Hamiltonian can be calculated by using $U=\exp{\left(-i \bar{\tilde{\tilde{\mathcal{H}}}}^{(1)} t\right)}$:

\begin{align}
	&\braket{I_{z}}(t) = 
	\frac{\gamma_{e}}{\gamma_{n}}\braket{\rho_{0}|\tilde{S}_{z}} Tr\{U \tilde{S}_{z} U^{\dagger} I_z\}/Tr\{I_z^2\} \nonumber \\ 
	&=	\pm
	\frac{\gamma_{e}}{\gamma_{n}}\braket{\rho_{0}|\tilde{S}_{z}} \frac{B^{2}a_{\mp}^{2}}
	{4 \omega_{\mp}^{2}} 
	\sin^{2}\left(\frac{1}{2} \omega_{\mp} t\right) \quad \text{for} \quad \omega_\text{eff}^{(S)}\approx\mp\omega_\text{eff}^{(I)} 
	\label{Eq:FOM}
\end{align}
with
\begin{align}
	a_{\mp} &= \sqrt{a_{-z}^{(\mp k_I)} a_{+z}^{(\pm k_I)}}
\end{align}

representing a unitless scaling factor that dictates the transfer efficiency. The quantity
\begin{align}
	\omega_{\mp} &=\sqrt{ B^{2}a_{\mp}^2/4+(\Delta\omega_\text{eff}^{\mp})^{2}}
\end{align}
characterizes the DNP buildup rate, and it depends on the ``mismatch'' of the effective fields:
\begin{align}
	\Delta\omega_\text{eff}^{\mp}(\Omega_S)&=	\omega_\text{eff}^{(S)}(\Omega_S) 
	\pm \omega_\text{eff}^{(I)} \quad,
	\label{eq:mismatch}
\end{align}
where the mismatch can be electron offset-dependent $\Delta\omega_\text{eff}^{\mp}(\Omega_S)$. While the buildup is faster in case of a larger mismatch, the transfer amplitude is lower, similar to the situation of an off-resonance pulse in a two-level system.
Note that the subscripts '-' and '+' in $a_{\mp}$ and $\omega_{\mp}$ symbolize  the ZQ and the DQ case, respectively. The prefactor 
$\frac{\gamma_{e}}{\gamma_{n}}\braket{\rho_{0}|\tilde{S}_{z}}$ in Eq. \ref{Eq:FOM} highlights that only the part of the electron density operator projected onto the effective field will be transferred to the nucleus via the ZQ/DQ operator. 

The ZQ/DQ-operator-mediated transfers may be visualized by expressing the Hamiltonian (Eq. \ref{eq:truncated first order ZQ}) in terms of fictitious spin-1/2 operators \cite{Vega1978,Wokaun1977} $I_x^\pm = \frac{1}{2}(\tilde{S}^+ I^\pm + \tilde{S}^- I^\mp)$, $I_y^\pm = \frac{1}{2i}(\tilde{S}^+ I^\pm - \tilde{S}^- I^\mp)$, and $I_z^\pm = \frac{1}{2}(\tilde{S}_z \pm I_z)$. This results in 
\begin{align}
	\bar{\tilde{\tilde{\mathcal{H}}}}^{(1)} =& 
	\frac{B}{2}\left(    \operatorname{Re}(a_{-z}^{(\mp k_I)}) I_x^\mp \mp \operatorname{Im}(a_{-z}^{(\mp k_I)}) I_y^\mp \right) \nonumber \\
	 +& 
	\Delta\omega_\text{eff}^- I_z^- +  \Delta\omega_\text{eff}^+ I_z^+
	\quad ,
	\label{eq:truncated first order ZQ fictitious}
\end{align}
recalling the superscript signs - and + relate to ZQ and DQ operators, respectively. Figure 1 shows the evolution of spin operators in a diabatic ('sudden') (Fig. 1a) or adiabatic (Fig. 1b) 
manner in the ZQ/DQ subspace. At the exact resonance condition (mismatch $\Delta\omega_\text{eff}^{\mp}=0$), the polarization-transfer expression simplifies to
\begin{equation}
	\braket{I_{z}}(t) = 
	\pm
	\frac{\gamma_{e}}{\gamma_{n}}\braket{\rho_{0}|\tilde{S}_z} \sin^{2}\left(\frac{B}{4} a_{\mp} t\right) \,\, \text{for} \,\, 
	\omega_\text{eff}^{(S)} = \mp\omega_\text{eff}^{(I)} \quad ,
	\label{eq:fom}
\end{equation}
whose initial polarization buildup (small $t$) can be approximated by a Taylor series as
\begin{equation}
	\braket{I_z}(t)  \approx 
	\pm
	\frac{\gamma_{e}}{\gamma_{n}}\braket{\rho_{0}|\tilde{S}_z} 
	\frac{B^2}{16}a_{\mp}^2t^2 + \mathcal{O}(t^4)  \,\, \text{for} \,\, 
	\omega_\text{eff}^{(S)} = \mp\omega_\text{eff}^{(I)} \quad .
	\label{eq:initial_dnp}
\end{equation}
We define the transfer parameter 

\begin{equation}
	f_{\mp} =	
	\braket{\rho_{0}|\tilde{S}_z} a_{\mp} \quad,
	\label{eq:transfer_coeff}
\end{equation}
which allows us to semi-quantitatively evaluate the performance of a pulse sequence, before performing a detailed analysis by defining a particular spin system. 

Finally, we should be aware that the choice of the effective field is not unique, and it requires a convention. We choose
$|\omega_\text{eff}^{(S)}|,|\omega_\text{eff}^{(I)}|\leq\omega_\text{m}/2$, which is an arbitrary but convenient decision. 
If we would allow for a larger effective field, it would become harder to keep track of resonance conditions. For our choice, there is one special case,
when $|\omega_\text{eff}^{(I)}|\approx|\omega_\text{eff}^{(S)}|\approx|\omega_\text{m}|/2$. 
In this case, the ZQ and DQ resonance conditions are approximately fulfilled at the same time, because the difference or sum of the effective 
field matches the modulation frequency implying that a single scaling factor is not sufficient to describe the spin dynamics.

\begin{figure}[htp]
	\centering
	\includegraphics[width=8.3cm]{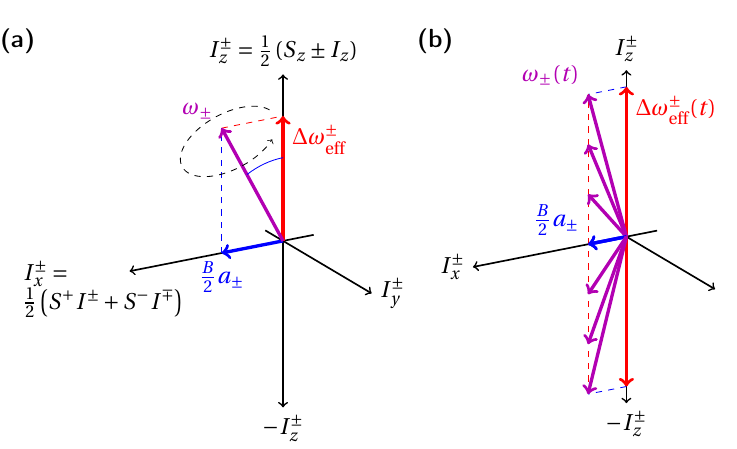}
	\vspace*{-0mm}
	\caption[]{Diagram of the effective (fictitious) spin-1/2 subspaces involved in DNP, with the - and + signs on operators and frequencies representing ZQ and DQ transfers, respectively. \textbf{(a)} A diabatic ('sudden') sequence with a constant mismatch. \textbf{(b)} Adiabatic sequence with slowly changing mismatch of the nuclear and electron effective fields. The effective ''mismatch'' field is slowly dragged from $+z$ to $-z$, corresponding to full polarization transfer. Note that $a_\mp$ is not necessarily a constant. In principle there can also be terms proportional to $I_y^\pm$, which we ignored for the illustration.
	}
	\label{fig:ZQ-subspace}
\end{figure}

\subsection{Illustration of the interaction frame transformation}

\begin{figure}[htp]
	\centering
	\includegraphics[width=8.3cm]{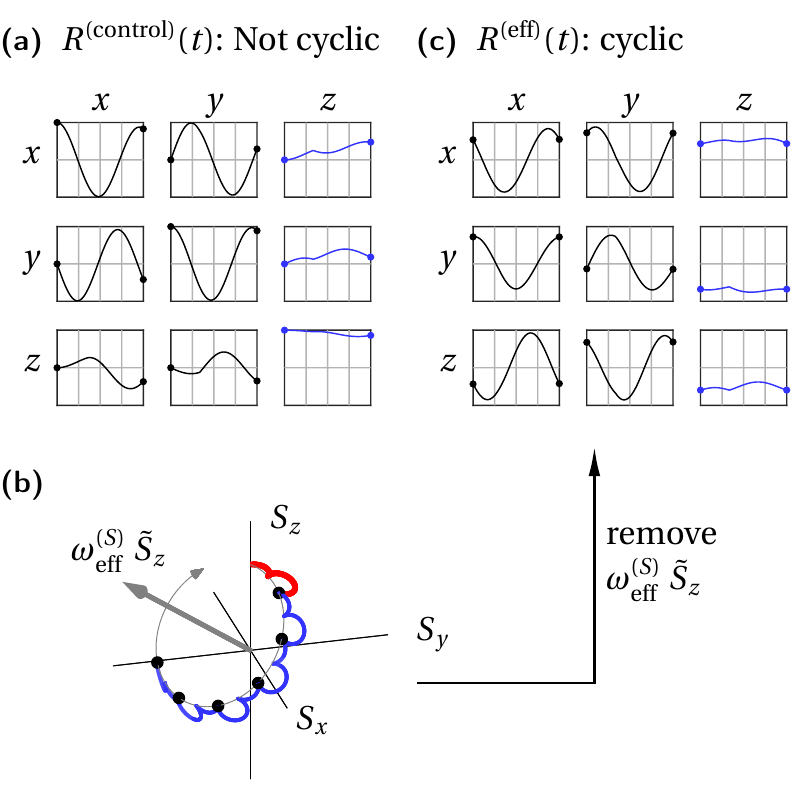}
	\vspace*{-0mm}
	\caption[DNP IFT Illustration]{Illustration for the interaction frames employed in this work, on the example of XiX-DNP (details in main text). \textbf{(a)} Illustration of the initial interaction frame transformation $R^{(\text{control})}(t)$. The individual plots show the evolution of rotation matrix elements over one period $\tau_\text{m}$. \textbf{(b)} Three-dimensional trajectory of the original $S_z$ operator in the initial interaction frame (the three blue components in \textbf{(a)}). The trajectory of the first modulation period is marked in red. The end points of the subsequent five periods are shown as black dots. The effective field describes the overall rotation of the sequence. \textbf{(c)} The same trajectory as in \textbf{(a)}, but in the flipped, effective (or cyclic) frame. The start and end points are the same.
	}
	\label{fig:IFT}
\end{figure}

A central element in calculating the scaling factors $a_{\mp}$ is the interaction frame transformations involved in Eqs. 
\ref{eq:rot_frame_ham2} - \ref{eq:rot_frame_ham3b}. Here, we give a concrete example of the elements of $R^{(\text{control})}$ and $R^{(\text{eff})}$ and their relationship.  We consider  an XiX-DNP experiment (parameters here not chosen to 
represent any good DNP sequence, but for easier explanation) with $\nu_1$=\,\unit[4]{MHz}, $t_{\text{p},1}$=\,\unit[14]{ns}, $t_{\text{p},2}$=\,\unit[28]{ns}, 
and $\Omega_S/2\pi$=\,\unit[25]{MHz} (see \autoref{fig:dnp sequences}
for the pulse sequence description).  
\autoref{fig:IFT}\textbf{(a)} shows the elements of the initial interaction frame transformation $R^{(\text{control})}(t)$ plotted over one period $\tau_\text{m}$. The blue curves denote the trajectory of the normal rotating frame operator $S_z$. Note that $R^{(\text{control})}(0)=\mathbb{1}\neq R^{(\text{control})}(\tau_\text{m})$, 
i.e. the trajectory is not cyclic with $\tau_\text{m}$. This prohibits the straight-forward application of average Hamiltonian theory and is the reason for the subsequent transformations. \autoref{fig:IFT}\textbf{(b)} shows the three-dimensional trajectory of the original $S_z$ 
operator in the initial interaction frame (the three blue components in panel \textbf{(a)}). The trajectory of the first modulation period is marked in red. 
The end points of the subsequent five periods are shown as black dots in panel \textbf{(b)}. The overall rotation from one period to the next can be described 
by an effective field shown in gray. This can be understood as a constant effective field, which can be removed by flipping the frame such that the effective field 
is along $z$, and then going into an interaction frame with said effective field. The result of this transformation, $R^{(\text{eff})}$, is shown 
in \autoref{fig:IFT}\textbf{(c)}. The $z$-axis in this new frame points along the effective field in \autoref{fig:IFT}\textbf{(b)}, and 
the effect of the overall rotation was eliminated by a counter rotation, i.e. start and end points of the trajectories are now the same.  
In \autoref{fig:IFT}\textbf{(c)}, all the coefficients are cyclic with time $\tau_\text{m}$. A Fourier transform of the respective time-dependent coefficients directly yields $a^{(k)}_{\chi z}$ in \autoref{eq:rot_frame_ham3b}.
An example script to perform these calculations in MATLAB is given in the SI.

\subsection{Adiabatic sweeps}

It is advantageous to  sweep the effective fields  across the resonance conditions because it can improve the  bandwidth and overall robustness of pulse sequences. Hence, we will address the effect of adiabatic sweeps through a resonance condition using the theoretical framework developed in this manuscript. For simplicity, we assume (1) only one resonance condition is swept during the experiment; (2) the changes in the scaling factors and the electron spin component along the effective field are sufficiently slow relative to the change in the effective fields, i.e., an adiabatic process.

To adiabatically invert the fictitious operator $I_z^\mp \rightarrow -I_z^\mp$ in the respective ZQ/DQ subspace (Eq. \ref{eq:truncated first order ZQ fictitious}) , an adiabatic sweep is implemented by varying the offset (or resonance mismatch) in the ZQ/DQ subspace starting from large positive values ($\Delta\omega_\text{eff}^{\mp} \gg0$), then slowly through zero ($\Delta\omega_\text{eff}^{\mp}=0$), and then continue to large negative values ($\Delta\omega_\text{eff}^{\mp} \ll0$). Note that the offset term in the ZQ/DQ subspace is usually not equivalent to the electron offset $\Omega_S$. For instance, in the adiabatic NOVEL DNP sequence, the offset in the ZQ/DQ subspace is determined by the mismatched Rabi field $\omega_{1S}(t)$, although the electron offset is $\Omega_S=0$ throughout the sequence. \autoref{fig:ZQ-subspace}\textbf{(b)} shows the schematic diagram of the described adiabatic sweep, which resulted in the spin evolution of $I_z^\mp \rightarrow I_z^\pm$ (in a subspace), which is mathematically equivalent to a $\tilde{S}_z \rightarrow I_z$ transfer.

To ensure that the sequence is adiabatic, one can calculate the adiabaticity $Q_\text{crit}$ at the moment the resonance condition is passed \cite{Baum1985,Jeschke2015}
\begin{align}
	Q_\text{crit}^\mp=\frac{1}{4}\frac{\left(B a_\mp  \right)^2}{\frac{\deriv}{\deriv t}\Delta\omega_{\text{eff}}^\mp(t)}
\end{align}
which can be exploited to evaluate the polarization-transfer efficiency using the Landau-Zener formula 
\begin{align}
	\braket{I_z} =\pm \frac{\gamma_{e}}{\gamma_{n}}\braket{\rho_{0}|\tilde{S}_{z}}\left(1-\exp\left( -\frac{\pi}{2} Q_\text{crit}^\mp\right) \right) \qquad. 
	\label{eq:Landau-Zener}
\end{align}
We note that it is impractical to vary the $\omega_{\text{eff}}$ indefinitely slow to maintain the high adiabaticity because relaxation effects  will start impeding the transfers at long mixing times, i.e., each sample/experiment has to be individually optimized for maximum transfer. Nevertheless, the scaling factor $a_\mp$ and the resonance conditions depend only on the pulse sequence, such that good initial guesses and sequence parameter ranges can be estimated theoretically.

\section{Materials and Methods}

\subsection{Numerical calculation of scaling factors}
All numerical calculations were implemented in MATLAB (The MathWorks Inc). All sequences presented in this work are piece-wise constant pulse sequences such that the effective fields could be calculated by quaternion multiplication of the individual pieces. Interaction-frame trajectories were calculated by time slicing. Fourier coefficients were calculated with an \texttt{fft} of $R^\text{(eff)}(t)$. The two-dimensional simulation of BASE-DNP was implemented with the simulation package SPINACH \cite{Hogben2011}. A three-spin electron-proton-proton system was used, with a $\mathbf{g}$-tensor of [2.0046 2.0038 2.0030], e-n distances of $r_1$=\,\unit[4.5]{\AA} and $r_1$=\,\unit[6.5]{\AA}, polar angles of $\theta_1$=$0^\circ$ and $\theta_2$=$90^\circ$ and azimuthal angles $\phi_1$=$0^\circ$ and $\phi_2$=$70^\circ$. A simple $T_1/T_2$ relaxation theory was used, with $T_{1,\text{e}}$=\,\unit[2.5]{ms}, $T_{2,\text{e}}$=\,\unit[5]{$\upmu$s} and $T_{1,\text{n}}$=\,\unit[36]{s}, $T_{2,\text{n}}$=\,\unit[1]{ms}. Relaxation was implemented via the Levitt-Di Bari approach\cite{Levitt1992}. A two-angle Lebedev grid\cite{Lebedev1999}  with 194 orientations was used. 

\subsection{Sample preparation}
A 5 mM sample of OX063 trityl radical in DNP juice (glycerol-d$_8$:\ce{D_2O}:\ce{H_2O}, 6:3:1 by volume) at 80 K was used for all experiments. In detail, \unit[1.65]{mg} trityl radical (MW=\,\unit[1359]{g mol$^{-1}$}, \unit[1.2]{$\upmu$mole}) were dissolved in \unit[24.3]{$\upmu$L} of \ce{H_2O} and \unit[72.9]{$\upmu$L} \ce{D_2O}. Of the resulting solution, \unit[48.6]{$\upmu$L} were then added to \unit[72.9]{$\upmu$L} of gly-$d_8$. \unit[40]{$\upmu$L} of the final solution were transferred to a \unit[3]{mm} OD quartz capillary and flash frozen in liquid nitrogen before the measurements.

\subsection{Instrumentation and EPR/NMR spectroscopy }
All experimental data were acquired on a new home-built X-band spectrometer which is based on the design described in Ref.(\cite{Doll2017}). Notable differences for the experiments described in this work were that a 1.8 GSa/s digitizer (SP Devices ADQ412) was used and that the temperature of \unit[80]{K} was achieved with a cryogen-free cryostat (Cryogenic Limited). Microwave pulses were generated with an arbitrary waveform generator (AWG) model  M8190A (Keysight) and amplified with a 1 kW traveling wave tube (TWT) amplifier (Applied Systems Engineering). A standard Bruker EN4118A-MD4 ENDOR resonator was used, with an external rf tuning and matching circuit.  NMR experiments were performed using a Stelar PC-NMR spectrometer. An Arduino board was used to count TWT gate triggers of the EPR spectrometer, each corresponding to an $h$ increment (Fig. \ref{fig:dnp sequences}), and the Arduino board  triggers the NMR acquisition after $h$ loops.  

FT EPR spectra were acquired by a chirp echo sequence with linear chirp pulses spanning  \unit[300]{MHz}, a duration of \unit[200]{ns} ($\pi/2$) and \unit[100]{ns} ($\pi$) and an inter-pulse delay of \unit[2]{$\upmu$s}. All EPR and NMR signals were processed in MATLAB. All  experimental results presented in this work were acquired within a single session, i.e., the sample was not moved between different DNP experiments.

\subsection{Pulse sequences and enhancements}

\begin{figure}[htp]
	\centering
	\includegraphics[width=\linewidth]{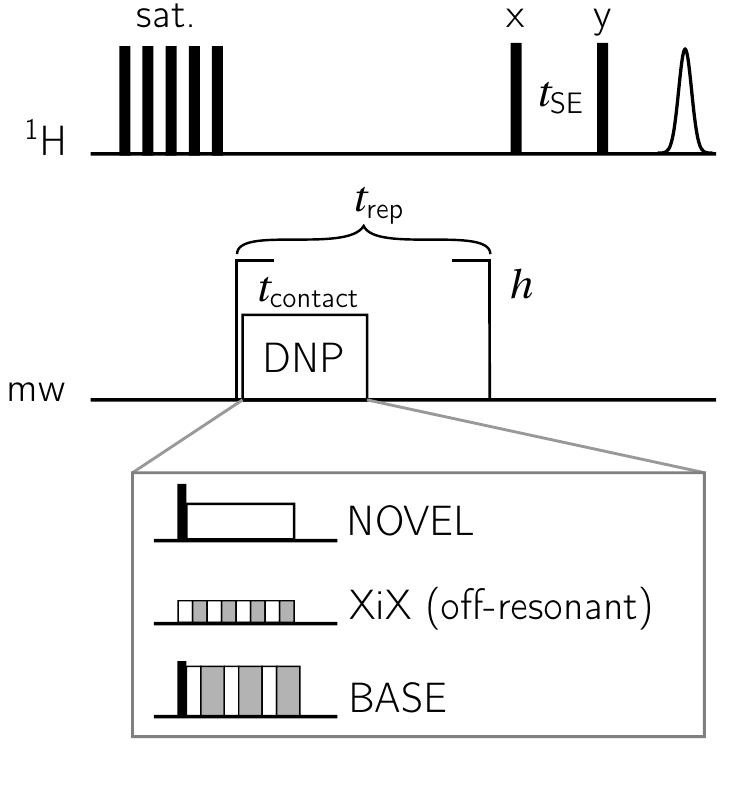}
	\caption[DNP pulse sequence]{General pulsed DNP sequence with various building blocks (NOVEL, XiX, and BASE) repeated by $h$ loops before $^1$H-NMR detection with a solid echo.
	}
	\label{fig:dnp sequences}
\end{figure}
The basic structure of all DNP experiments, as addressed individually in the following section, is shown in \autoref{fig:dnp sequences}.  A train of $^{1}$H saturation pulses  (eleven 100$^\circ$ pulses spaced by \unit[1]{ms}) was applied before the DNP element. Each DNP block was repeated $h$ times, with a total build-up time $T_\text{DNP}=h\cdot t_\text{rep}$. The contact time $t_\text{contact}$, during which the microwaves are turned on, is generally much shorter than the repetition time $t_\text{rep}$,  due to a 1\% duty cycle limit of the TWT. The $^1$H NMR signal was then read out with a solid echo sequence comprised of two  \unit[2.5]{$\upmu$s} $90^\circ$ pulses separated by a delay of $t_\text{SE}$=\,\unit[80]{$\upmu$s}. A conventional eight-step phase cycle was used with $\{x,x,y,y,-x,-x,-y,-y\}$ for the first pulse and detection and $\{y,-y,x,-x,y,-y,x,-x\}$ for the second pulse. The  proton spectrum at thermal equilibrium was acquired using similar parameters except without microwaves, and a delay of   \unit[180]{s} $\approx 5\cdot T_{1,\text{n}}$  was used in between the  660 accumulated scans. The $T_{1,\text{n}}$=\,\unit[36]{s} was determined both with a saturation recovery sequence, and by the decay of polarization after DNP (see SI).

For most cases, we report the \textit{polarization enhancement} $\varepsilon_P$, given by the ratio of the DNP-enhanced signal intensity divided by the signal  intensity at thermal equilibrium. These values can be different from simple mw on/off signal enhancements recorded with the same delay, because the DNP build-up time $T_B$ can be much shorter than $T_{1,\text{n}}$. For most parameter optimizations, we used a repetition time $t_\text{rep}$ of \unit[1]{ms}, and a build-up time $T_\text{DNP}$ of \unit[2]{s}. Build-up curves were acquired by changing the value of $h$, and with variable repetition times mentioned in the respective figures.

\section{Results}

In this section, we apply the theory and design procedures outlined earlier for DNP experiments at X-band frequencies (\unit[9.5]{GHz}) using both low-power and high-power microwaves. The former involves variants of the recently published XiX-DNP experiments \cite{Mathies2022}, while the latter involves development of a new pulse sequence with improved performance relative to previous NOVEL-DNP experiments \cite{Henstra1988,Can2015a,Can2017b}.

\subsection{Low-power XiX-DNP}

\begin{figure*}[htp]
	\centering
	\includegraphics[width=0.9\linewidth]{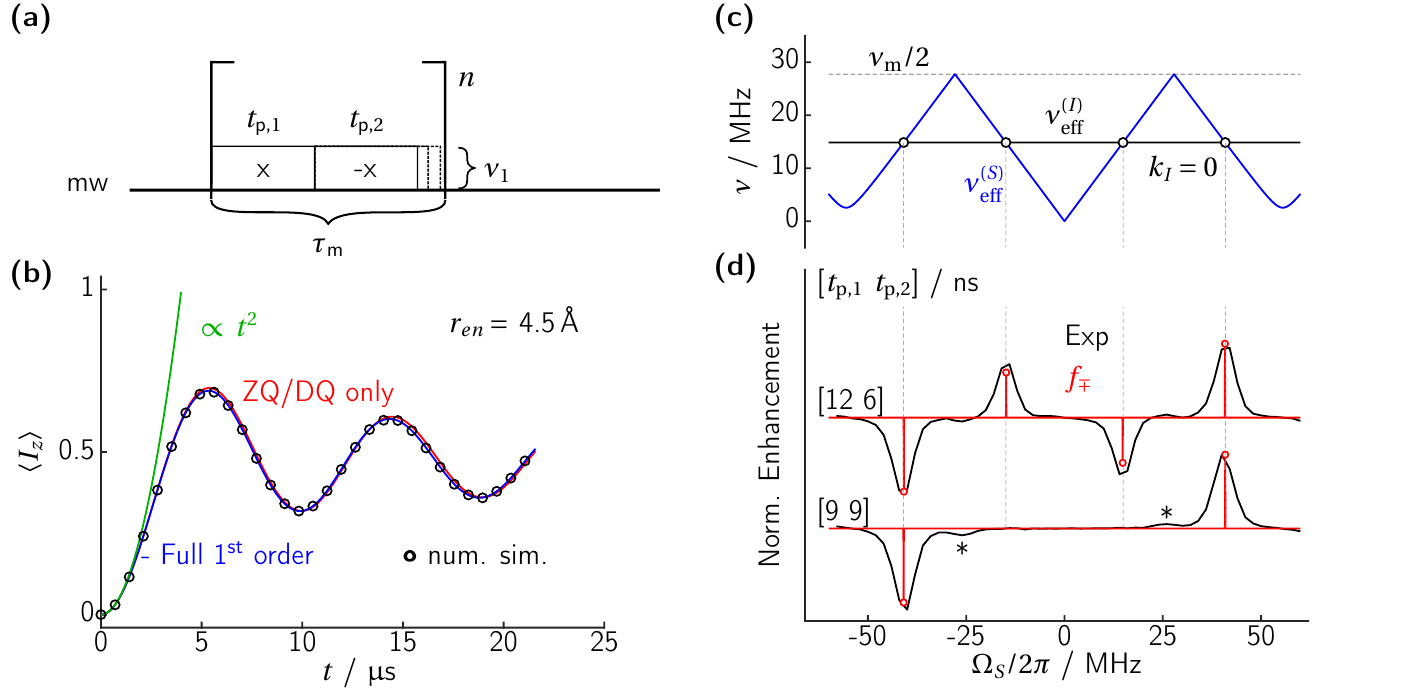}
	\vspace*{-0mm}
	\caption[]{
		Numerical and experimental analysis of the XiX-DNP experiment. \textbf{(a)} Pulse sequence for the mw part of the DNP experiment. \textbf{(b)} Comparison of $\tilde{S}_z \rightarrow I_z$ polarization transfer efficiencies calculated using an effective Hamiltonian  including all first order terms (blue) or only the flip-flop terms (red) 
		with a full numerical simulation (black circles). A two-spin e-$^1$H spin pair with a distance $r_{en}$ = 4.5 Å is used in the numerical simulations. 	The green line illustrates the initial build-up in \autoref{eq:initial_dnp}. \textbf{(c)} and \textbf{(d)} Resonance conditions and theoretical and experimental enhancements for XiX-DNP with $\nu_1$=\,\unit[4]{MHz},  $t_\text{contact}$=\,\unit[8]{$\upmu$s} $\tau_\text{rep}$=\,\unit[1]{ms}, $T_\text{DNP}$=\,\unit[2]{s} as function of the electron offset frequency. \textbf{(c)} The absolute value of the effective fields $\nu_\text{eff}^{(S)}$ (blue) and $\nu_\text{eff}^{(I)}$ (black) as a function of the mw offset, for $t_{\text{p},1}=t_{\text{p},2}=$\,\unit[9]{ns}. Resonance conditions are indicated as black circles. \textbf{(d)}  Experimental enhancements for different combinations of $t_{\text{p},1}$ and $t_{\text{p},2}$ (given in brackets) with fixed total modulation period (black), and theoretical predictions based on \autoref{eq:transfer_coeff} (red). Small additional peaks are due to higher-order processes involving two protons. The calculation in \textbf{(b)} was done at an electron offset of \unit[40.89]{MHz}. 
	}
	\label{fig:XiX1}
\end{figure*}

 \autoref{fig:XiX1}\textbf{(a)}
shows the mw part of the XiX-DNP pulse sequence consisting of two oppositely phased pulses repeated $n$ times, leading to a total contact time of $t_\text{contact}=n\cdot\tau_\text{m}=n\cdot(t_{\text{p},1}+t_{\text{p},2})$. Assuming an mw field with an amplitude of $\nu_1$=\,\unit[4]{MHz} (we use $\omega$ for angular frequencies and $\nu=\omega/2\pi$ for linear frequencies), and an offset slightly above \unit[40]{MHz}, $t_{p,1}$ = $t_{p,2}$ = 9 ns, this leads to the calculated transfer profiles shown in  \autoref{fig:XiX1}\textbf{(b)} when using fully numerical simulations (black circles), the full first-order Hamiltonian (blue), partial ZQ/DQ Hamiltonians (Eq. \ref{eq:truncated first order ZQ}, red), or Taylor-expanded series (Eq. \ref{eq:initial_dnp}, green). The offset-dependent resonance conditions become clear in \autoref{fig:XiX1}\textbf{(c)}
showing the electron and nuclear effective fields and the matching conditions. In this example, $\nu_\text{eff}^{(I)}=\nu_I$ and $k_I=0$ for all resonance conditions. Since low-power mw irradiation is used, the electron effective field is mainly dominated by the electron offset. We note the reflection at $\nu_\text{m}/2$, which is a consequence of our particular choice of convention. 

\autoref{fig:XiX1}\textbf{(d)} shows the experimental results and calculated $f_\mp$ (\autoref{eq:transfer_coeff}) for different combinations of $t_{\text{p},1}$ and $t_{\text{p},2}$ (but with a constant sum $t_{\text{p},1}+t_{\text{p},2}$).    The bottom case with $t_{\text{p},1}=t_{\text{p},2}$ corresponds to the sequence introduced by Mathies et al.\cite{Mathies2022}.  Clearly, both the positions and the relative intensities of the matching conditions are well predicted. The small peaks visible in the experimental data correspond to a three-spin electron-$^1$H-$^1$H  transition (see SI). Interestingly,  if both pulses have the same length (the bottom trace in \autoref{fig:XiX1}\textbf{(d)}), the resonance condition at the usual SE offset ($\Omega_S/2\pi\approx\nu_I$) is still fulfilled, but the scaling factor is zero. 
This figure shows that the resonance conditions alone are not enough to characterize the DNP performance and that the theoretical scaling factors reliably predict the relative DNP enhancement.

\begin{figure*}[htp]
	\centering
	\includegraphics[width=\linewidth]{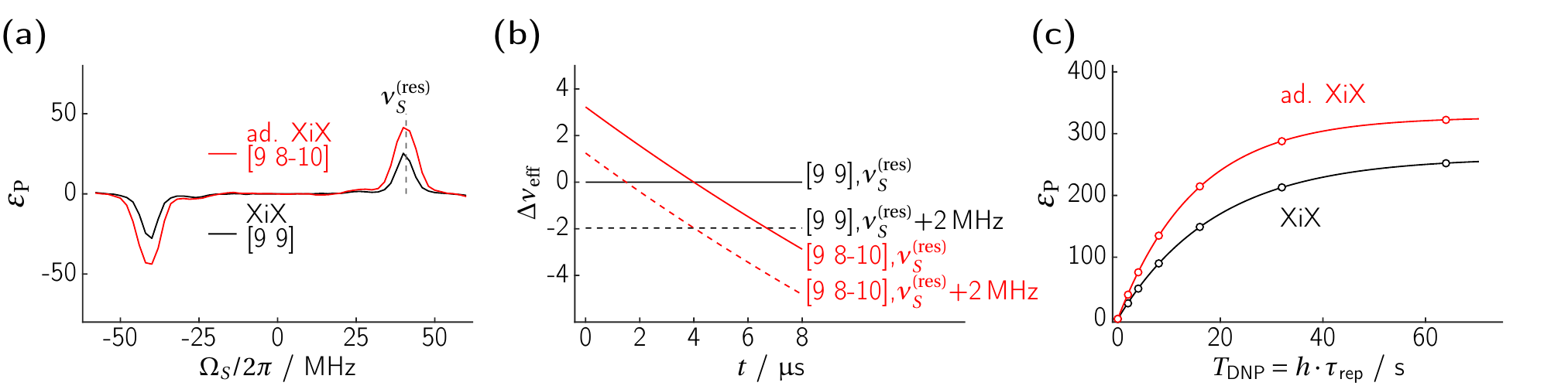}
	\vspace*{-0mm}
	\caption[]{		
		Experimental comparison of the diabatic ($t_{p,1}$=$t_{p,2}$=\,\unit[9]{ns}, black) and adiabatic  ($t_{p,1}$=\,\unit[9]{ns}, $t_{p,2}$=\,\unit[8-10]{ns}, red) XiX-DNP.	\textbf{(a)} XiX-DNP mw offset profile with \unit[2]{s} of buildup time. \textbf{(b)} Mismatch between nuclear and electron effective fields for XiX (black) and its adiabatic version (red) as a function of contact time.
		Solid lines show an exactly-matched ($\Delta\nu_{\textrm{eff}}=0$) resonance condition around \unit[40]{MHz}, whereas the dashed line describes a scenario of a shifted resonance condition by 2 MHz. All lines except the black dashed line cross the $\Delta\nu_{\textrm{eff}}=0$ line, and hence DNP will take place. This clearly shows the mismatch compensating feature exhibited by adiabatic sequences. \textbf{(c)} Experimental $^1$H build-up curves  with a repetition time of $t_\mathrm{rep}=$\,\unit[1]{ms}. XiX-DNP: $\varepsilon_\mathrm{max}=$\,261, $T_\mathrm{B}=$\,\unit[19.0]{s}, adiabatic XiX-DNP: $\varepsilon_\mathrm{max}=$\,327, $T_\mathrm{B}=$\,\unit[15.1]{s} }
	\label{fig:adXiX}
\end{figure*}

 The performance of the XiX-DNP experiment may be further improved  by adiabatically sweeping the effective fields through the resonance condition, i.e., slowly increase $t_{\text{p},2}$ upon increasing the loop number $n$ (\autoref{fig:XiX1}\textbf{(a)}). The improved enhancement is  demonstrated in DNP experiments (\autoref{fig:adXiX}\textbf{(a)}), where the adiabatic version of XiX-DNP with the second pulse swept from 8--10 ns (red line) clearly outperforms its diabatic counterpart proposed by Mathies and coworkers \cite{Mathies2022} (black line). \autoref{fig:adXiX}\textbf{(b)} shows the time-dependence of the effective field mismatch $\Delta\omega^-_{\text{eff}}(t)$ for the diabatic and adiabatic variants. The black lines correspond to the diabatic variant with fixed timing. Exactly  at the offset of \unit[40.89]{MHz}, the effective field mismatch is exactly zero (black solid line), and it does not change over time. Under these conditions, the transfer is optimal. However, if the offset is  \unit[2]{MHz} off --- a reasonable value given that the FWHM of trityl is $\sim$\unit[5]{MHz} --- the effective fields are also mismatched by about \unit[2]{MHz} (black dashed line). Consequently, the DNP matching condition is not fulfilled and the DNP transfer is entirely quenched for a system with small hyperfine couplings. On the contrary, it is evident that the effective field of the adiabatic variant (red) crosses zero in both cases, leading to polarization transfer for a broader distribution of electron offsets.
Finally, \autoref{fig:adXiX}\textbf{(c)} shows the experimental build-up curve of the two  XiX-DNP sequences, which clearly shows the advantages of implementing adiabatic XiX-DNP.

\subsection{High-power BASE-DNP}

\begin{figure*}[htp]
	\centering
	\includegraphics[width=0.9\linewidth]{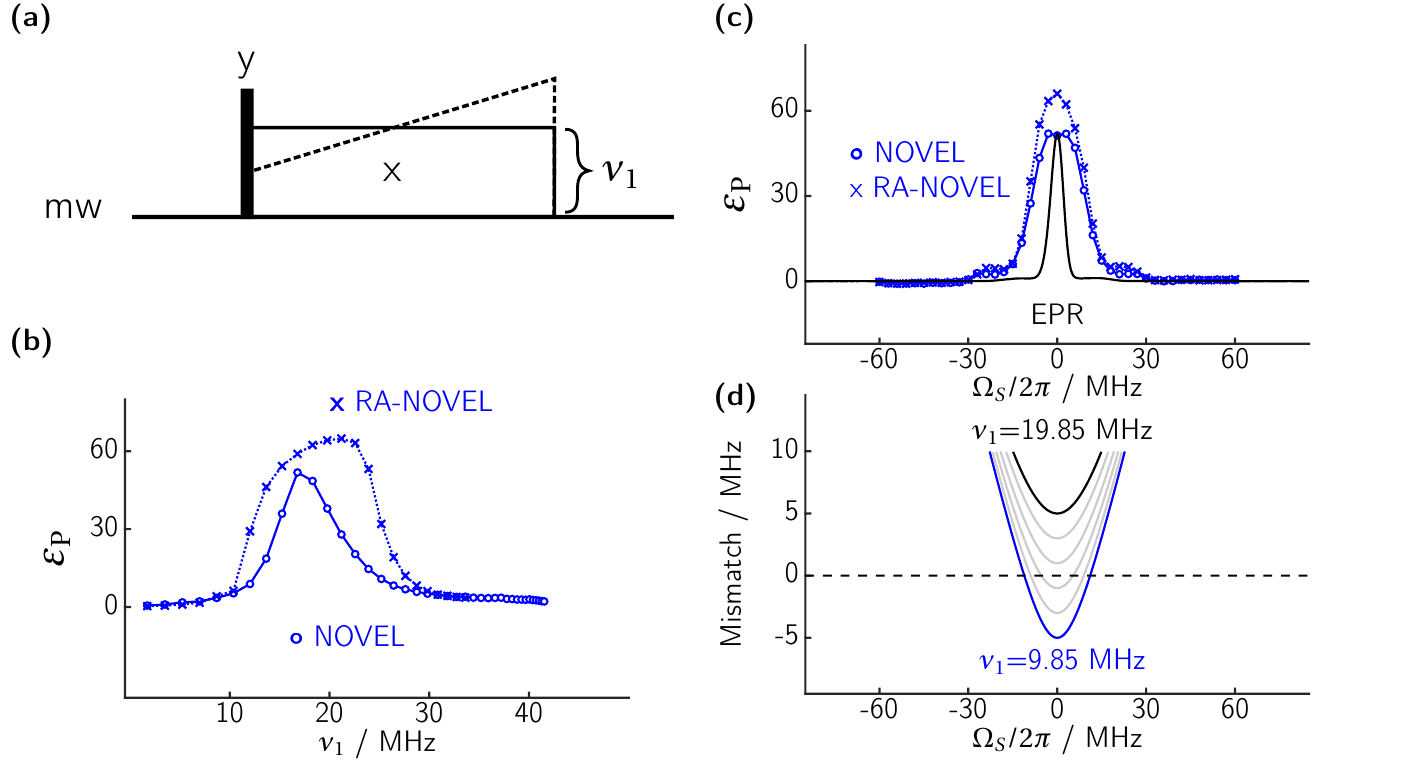}
	\vspace*{-0mm}
	\caption[]{ \textbf{(a)} Pulse sequence for NOVEL and Ramped Amplitude (RA) NOVEL (dashed) DNP. \textbf{b)} Experimental DNP enhancement as a function of the spin lock strength (Rabi frequency) $\nu_1$ after \unit[2]{s} of DNP. \textbf{(c)} DNP mw offset profiles for both NOVEL sequences (blue) and the EPR spectrum (black) with an arbitrary scale. \textbf{(d)} Calculated resonance mismatch (Eq. \ref{eq:mismatch}) as a function of the offset for RA-NOVEL. The adiabatic sequence begins  with a large negative $\Delta\nu_{\textrm{eff}}\ll0$ (blue), which slowly increases towards $\Delta\nu_{\textrm{eff}}\sim0$ (gray), and ends with a large positive $\Delta\nu_{\textrm{eff}}\gg0$ (black). DNP occurs whenever the lines cross $\Delta\nu_{\textrm{eff}}=0$. }
	\label{fig:RANOVEL}
\end{figure*}

We will now analyze pulsed DNP sequences requiring  high-power mw irradiation. In particular, we examine  NOVEL (nuclear orientation via spin locking)  \cite{Henstra1988,Can2015a,Mathies2016} and its adiabatic version, the ramped-amplitude RA-NOVEL-DNP \cite{Can2017b} (\autoref{fig:RANOVEL}\textbf{(a)}). We will then show how a simple amplitude modulation can be used to improve its bandwidth.

For NOVEL, the spinlock strength has to match the nuclear Zeeman frequency, $\nu_1\approx\nu_I$, while for RA-NOVEL the nutation frequency is slowly increased from below the matching condition to above it in a linear fashion. Although other amplitude modulation regimes were examined, no major improvement was observed at long contact times \cite{Can2017b}. \autoref{fig:RANOVEL}\textbf{(b)} and \textbf{(c)} compare the DNP performances of the NOVEL sequences as function of the (average) Rabi field  $\nu_1$ and the offset $\Omega_S/2\pi$. The plots shows that  RA-NOVEL is more tolerant towards $\nu_1$ mismatch, and hence leads to higher DNP enhancements. Additionally, the calculated mismatch plot (\autoref{fig:RANOVEL}\textbf{(d)}) also predicts that the adiabatic sequence can moderately improve the bandwidth for small couplings. Nevertheless, RA-NOVEL experiments did not show an improved offset compensation, most likely due to the more dominant mw Rabi field inhomogeneity (about \unit[18]{\%}) across the sample.

\begin{figure*}[htp]
	\centering
	\includegraphics[width=0.9\linewidth]{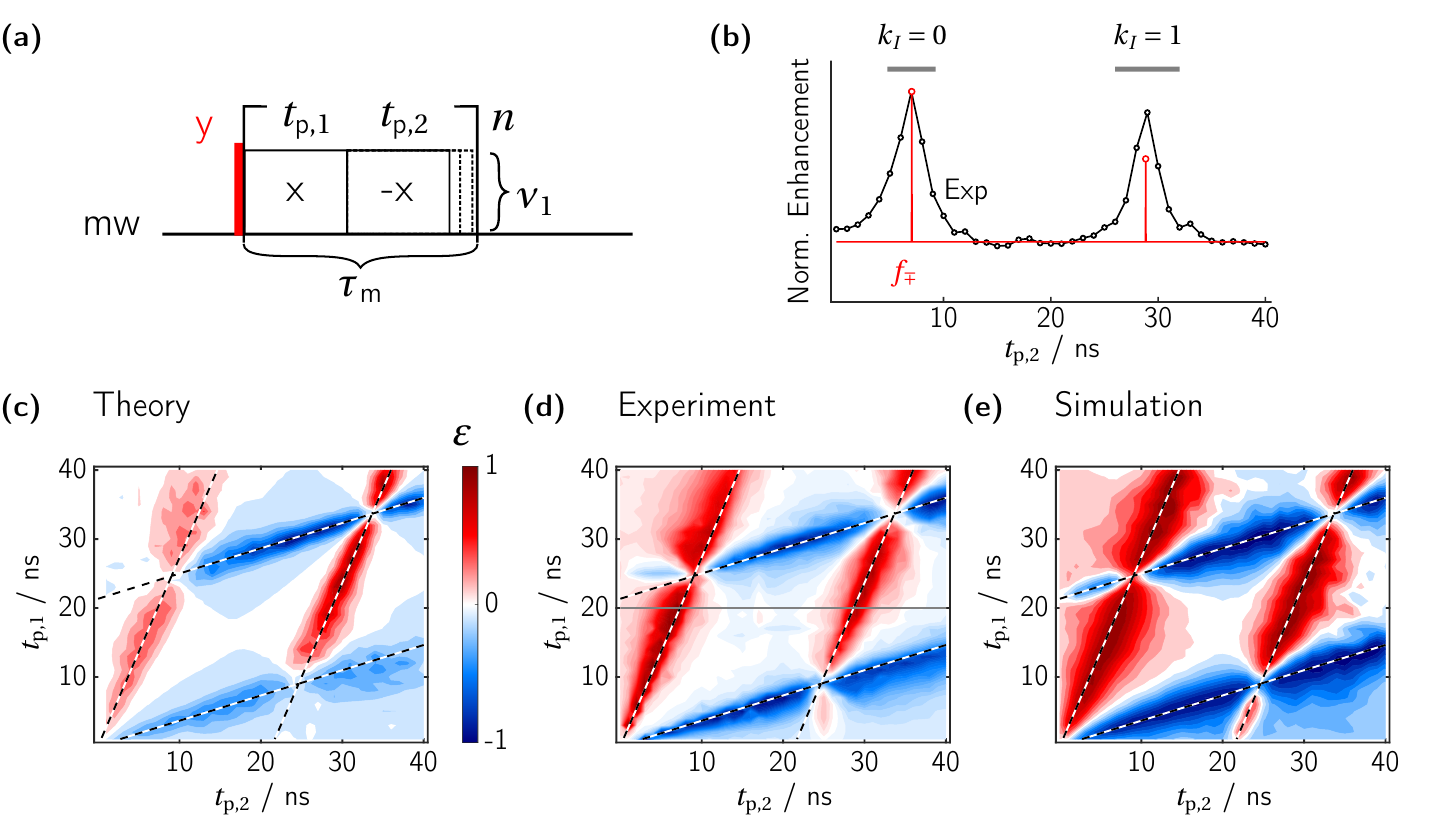}
	\vspace*{-3mm}
	\caption[]{ \textbf{(a)} Pulse sequence of BASE-DNP. \textbf{(b)} Experimental  (black) and calculated  (red) DNP performance as a function of $t_{\text{p},2}$ with fixed $t_{\text{p},1}$=\unit[20]{ns}. The ranges for the adiabatic sweeps are marked by gray bars. \textbf{(c)} Calculated relative enhancement (\autoref{Eq:FOM}) including offset distributions (\unit[5]{MHz} FWHM) and $\nu_1$  inhomogeneity (\unit[6]{MHz} FWHM centered at \unit[32]{MHz}).  \textbf{(d)} Experimental BASE-DNP enhancement as a function of $t_{\text{p},1}$ and $t_{\text{p},2}$ with  $T_\text{DNP}$=\,\unit[1]{s}. The observed resonance conditions matches well with the theory (black and white dashed lines). The solid gray line indicates the position of \textbf{(b)}. \textbf{(e)} SPINACH simulation on a spin system described in Materials and Methods.  The experiments were performed using on-resonance ($\Omega_{S}=0$) mw irradiation and a Rabi field of $\nu_1\sim$ 32 MHz, which is twice of that used for NOVEL. Other experimental details include $t_\text{contact}$=\,\unit[8]{$\upmu$s}, $\nu_1\approx$\,\unit[32]{MHz}, $t_\text{rep}$=\,\unit[1]{ms}.
	}
	\label{fig:BASE1}
\end{figure*}

Motivated by these results and also our previous works in designing broadband ssNMR recoupling sequences \cite{Tan2014}, we hypothesized that a broadband pulsed DNP sequence can be designed by combining XiX and NOVEL, i.e., a Broadband Amplitude modulated Signal Enhanced (BASE) DNP (\autoref{fig:BASE1}\textbf{(a)}). Similar to NOVEL (but unlike the XiX DNP), BASE-DNP is a spin-locked experiment. Additionally, the sequence can be made adiabatic by slowly varying  $t_{\text{p},2}$ through one of the matching conditions, which are shown as dashed lines in the 2D plot of a theoretical prediction using  Eq. \ref{eq:fom} (\autoref{fig:BASE1}\textbf{(c)}). This calculation included distributions of electron offsets and Rabi fields. The intensity and width of the resonance conditions already hint at the robustness of them with respect to these parameters.

\autoref{fig:BASE1}\textbf{(b)} shows the DNP enhancement as a function of $t_{\text{p},2}$ with a fixed value of $t_{\text{p},1}$=\,\unit[20]{ns}. One can see that the calculated $f_\mp$ (Eq. \ref{eq:transfer_coeff}) matches  the observed resonance conditions and the relative DNP performance  well, despite the fact that
the Rabi field inhomogeneity and mw offsets were simply neglected in these calculations. We have labeled the two different resonance conditions ($k_I=0$ and $k_I=1$) for later reference (\textit{vide infra}), and the sweep range of the adiabatic variants are indicated by gray bars.
The experimental enhancement as a function of both pulse lengths is shown in \autoref{fig:BASE1}\textbf{(d)}. Again the positions of the resonance conditions are well predicted by the theory. There are some differences in the width and intensity that are expected from the simplistic two-spin model we are using. A slightly more sophisticated (though still simplistic) numerical calculation employing Spinach \cite{Hogben2011} is shown in \autoref{fig:BASE1}\textbf{(e)}. 
Overall, our theory reliably predicts the resonance conditions for BASE-DNP. While numerical simulations of small spin systems can include more details, such as electronic relaxation, they are still not capturing all the complications in the complete DNP process. In this case, our (semi-)analytical theory is very helpful in quickly identifying resonance conditions and for choosing suitable experimental parameters.

\begin{figure}[htp]
	\centering
	\includegraphics[width=0.9\linewidth]{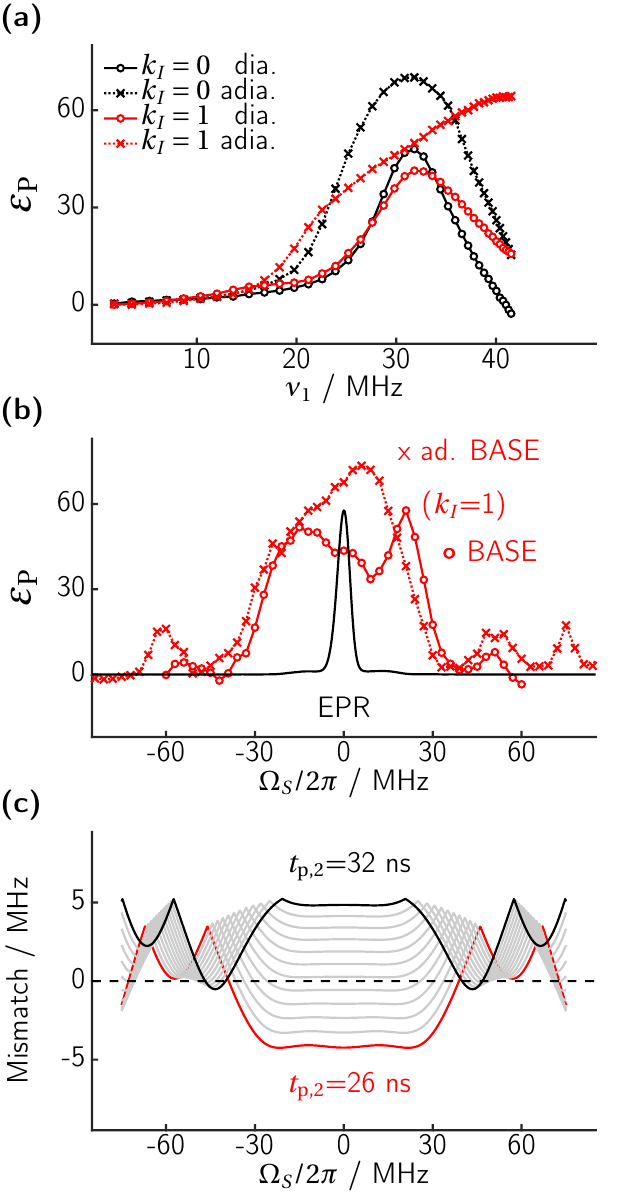}
	\vspace*{-3mm}
	\caption[]{Experimental BASE-DNP enhancement as a function of \textbf{(a)} the Rabi field $\nu_1$ for the diabatic and adiabatic versions of the respective resonance conditions and \textbf{(b)} offset $\Omega_S/2\pi$, and an EPR spectrum is included here for reference (black). The BASE parameters were $t_{\text{p},1}$=\,\unit[20]{ns}, $t_\text{contact}$=\,\unit[8]{$\upmu$s}, $\tau_\text{rep}$=\,\unit[1]{ms}, $T_\text{DNP}$=\,\unit[2]{s}. The pulse length of the second pulse, $t_{\text{p},2}$ was fixed in the case of (diabatic) BASE ('$\circ$') to \unit[7]{ns} ($k_I=0$) and \unit[29]{ns} ($k_I=1$), while it  was swept over {4.75}--{9.25} {ns} ($k_I=0$) or {26}--{32} {ns} ($k_I=1$) for adiabatic BASE ('x').  \textbf{(c)} Calculated offset-dependent mismatch  for  $k_I=1$ condition. The adiabatic sweep begins with $t_{\text{p},2}=$ \unit[26]{ns} (red), through the intermediate stages (gray), and ends at $t_{\text{p},2}=$ \unit[32]{ns} (black). 
			}
	\label{fig:BASE2}
\end{figure}

We will now characterize the BASE-DNP resonance conditions in more detail.
The DNP enhancement as a function of  $\nu_1$ is shown in \autoref{fig:BASE2}\textbf{(a)}.  Is is  evident that the adiabatic  BASE outperforms its diabatic counterpart. For one of the resonance conditions, the best transfer was achieved with the highest power available. A closer inspection at the $k_I=0$ and $k_I=1$ resonance conditions reveals that the position of the latter is much more robust with respect to the spin-lock field strength, in agreement with the experimental data (\autoref{fig:BASE2}\textbf{(a)}). \autoref{fig:BASE2}\textbf{(b)} shows the BASE DNP frequency profile (constant  $B_\textrm{0}$ field  with varying mw center frequency) for the $k_I=1$ case. A maximum mw power was used for the $\pi/2$ pulse for a maximum bandwidth, but the spin-lock field was adjusted at each offset position according to the mw resonator (see SI). Note that the mw power adjustment was not  possible for the adiabatic BASE due to the limited mw power available. The small enhancements at larger offsets ($\frac{\Omega_{S}}{2\pi}=\pm 60 $ MHz) are due to the matched resonance conditions during the adiabatic sweep, where the offset-dependent mismatch during the contact period is explicitly calculated (\autoref{fig:BASE2} \textbf{(c)}). 

\begin{figure}[htp]
	\centering
	\includegraphics[width=0.9\linewidth]{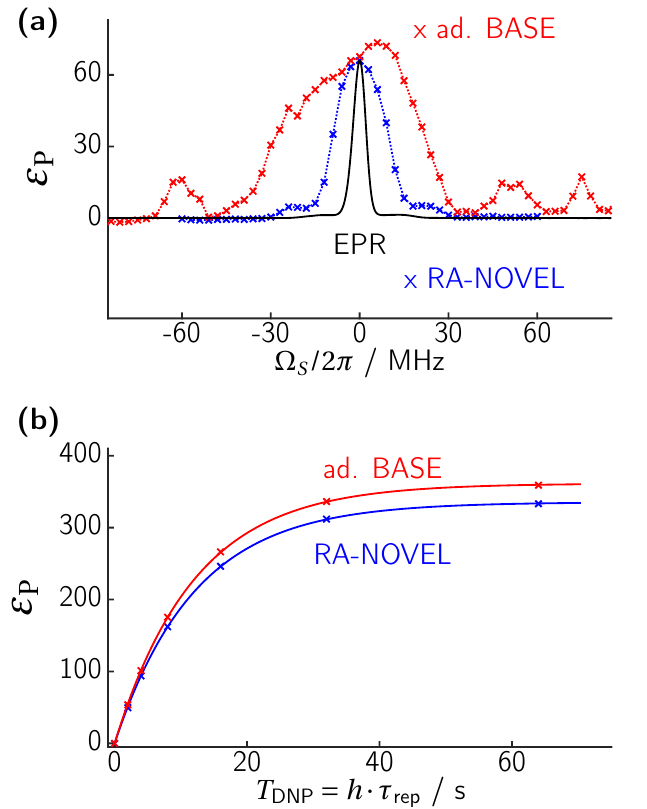}
	\vspace*{-3mm}
	\caption[]{  \textbf{(a)} DNP frequency profile of  RA-NOVEL and adiabatic BASE ($k_I=1$) DNP with \unit[2]{s} build up time.  \textbf{(b)} DNP build-up curve using a repetition time of \unit[5]{ms}. The build-up curves were fitted with exponential functions $\varepsilon_\text{P}(T_\text{DNP})=\varepsilon_\text{max}\left(1-\exp(-T_\text{DNP}/T_B) \right)$. The build-up curves were measured using different repetition times  $\tau_\text{rep}$ \unit[1-20]{ms}, where $t_\mathrm{rep}=$~5~ms yield the largest $\varepsilon$, and $t_\mathrm{rep}\sim$ 2~ms resulted in the highest $\varepsilon_\text{max}/\sqrt{T_{1,\text{n}}/T_B}$. $90^\circ$ flip-back pulses were applied after each DNP contact to replenish the electron Zeeman spin bath. Exact BASE parameters are given in \autoref{fig:BASE2}.
	}
	\label{fig:RANOVEL_vs_BASE}
\end{figure}

As discussed in the analysis performed for RA-NOVEL,  DNP occurs when the mismatch is zero (diabatic case) or passes zero (adiabatic case). The theory shows that the mismatch for BASE DNP is quite offset-tolerant, as visible in \autoref{fig:BASE2} \textbf{(c)}. In other words, the mismatch hardly varies by more than 1 MHz over the $|\Omega_{s}/ 2\pi| <$ 30 MHz range. Hence, the theory implies that BASE will be a broadband sequence, and indeed this was verified experimentally.
It is evident that the increased Rabi field $\nu_1$ employed in BASE has resulted in a $\sim3\times$ higher bandwidth compared to RA-NOVEL (\autoref{fig:RANOVEL_vs_BASE}).
 Moreover, we also measured  the build-up curves (\autoref{fig:RANOVEL_vs_BASE}\textbf{(b)}), and the results  are summarized in \autoref{tab:enhancements}. In summary, adiabatic BASE has outperformed RA-NOVEL with a higher $\varepsilon_\mathrm{max}$ of $\approx$ 361 and $\varepsilon_\mathrm{max}\cdot\sqrt{{T_\mathrm{1,n}}/{T_\mathrm{B}}}\approx$ 701, compared to $\varepsilon_\mathrm{max}$ of $\approx$ 335 and $\varepsilon_\mathrm{max}\cdot\sqrt{{T_\mathrm{1,n}}/{T_\mathrm{B}}}\approx$ 671 for RA-NOVEL. Although the improvement of adiabatic BASE over RA-NOVEL is marginal ($\leq 8\%$), a larger relative gain can be envisaged when applied to other more generic DNP polarizing agents---which usually have broader lines than OX063, and, hence, offset compensation becomes critical.

\begin{table}[htp]
	\ra{1.1}
	\caption{Enhancements $\varepsilon_\mathrm{max}$, build-up times $T_\mathrm{B}$ , and sensitivity per unit time (i.e., signal per square root of time) $\varepsilon_\mathrm{max}\cdot\sqrt{{T_\mathrm{1,n}}/{T_\mathrm{B}}}$ for RA-NOVEL and adiabatic BASE. All measured with an additional flip-back pulse after the DNP contact. $T_\mathrm{1,n}=$~36.2~s. $T_\mathrm{1,e}=$~2.5~ms.}
	\begin{ruledtabular}
	\begin{tabular}{l c c }
		& RA-NOVEL & ad. BASE\\
		\hline
		& \multicolumn{2}{c}{$t_\mathrm{rep}=$~2~ms} \\
		\cmidrule{2-3}
		$\varepsilon_\mathrm{max}$& 321 & 342 \\
		$T_\mathrm{B}$ / s& 8.3 & 8.7  \\
		$\varepsilon_\mathrm{max}\cdot\sqrt{\frac{T_\mathrm{1,n}}{T_\mathrm{B}}}$& 671 & 701 \\
		\hline
		& \multicolumn{2}{c}{$t_\mathrm{rep}=$~5~ms} \\
		\cmidrule{2-3}
		$\varepsilon_\mathrm{max}$& 335 & 361 \\
		$T_\mathrm{B}$ / s& 12.1 & 12.1  \\
		$\varepsilon_\mathrm{max}\cdot\sqrt{\frac{T_\mathrm{1,n}}{T_\mathrm{B}}}$& 580 & 626 \\
	\end{tabular}
\end{ruledtabular}
	\label{tab:enhancements}
\end{table}

\section{Conclusions and outlook}
We have demonstrated a generalized theoretical treatment applicable to periodic DNP sequences in static samples. This is realized by analyzing the resonance conditions and determining the Fourier coefficients, which encode the details in a sequence --- Rabi fields, phases, amplitudes, and mw offsets. We showed here an example of how theory can help design a broadband sequence that is robust against mw offsets. With further improvement by implementing adiabatic sweeps, we show that adiabatic BASE, to the best of our knowledge, is the best performing pulsed DNP sequence discovered to date. This is supported by having obtained excellent enhancement values on trityl at static conditions, a temperature of \unit[80]{K} and a field of \unit[0.35]{T} --- with $\varepsilon_\mathrm{max}$ of $\approx$ 361 and $\varepsilon_\mathrm{max}\cdot\sqrt{{T_\mathrm{1,n}}/{T_\mathrm{B}}}\approx$ 701, which are higher than RA-NOVEL. While the adiabatic solid  effect\cite{Tan2020} can achieve similar maximal enhancements of 360, it can only reach sensitivity enhancements of $\varepsilon_\mathrm{max}\cdot\sqrt{{T_\mathrm{1,n}}/{T_\mathrm{B}}}\approx$ 629 (see SI) and is inherently limited to a bandwidth equal to or lower than the nuclear Zeeman frequency.

Moreover, our theory implies that adiabatic BASE should have field-independent performances, provided that the mw Rabi fields are also scaled linearly with the external magnetic fields. Although the  mw power requirement for adiabatic BASE currently cannot be fulfilled for high-field ($>5$ T) DNP NMR applications, it could be satisfactorily fulfilled for polarizing \ce{^{13}C} in diamond nitrogen-vacancy (NV) centers. For example,  PulsePol, which requires an order of magnitude more mw power than adiabatic BASE, was demonstrated to polarize \ce{^{13}C} nuclei in diamond NVs at $\sim$ 0.17 T \cite{Schwartz2018a}. In such situations, a broadband sequence that is robust against mw power inhomogeneity and offsets could be  advantageous for quantum computing applications \cite{Scheuer2020}.

Lastly, we emphasize that our generalized theoretical framework is not only applicable to DNP sequences, it is also valid for other magnetic resonance applications. In fact, the framework developed in this paper is a direct adaptation (with minor adjustments) from NMR theory, as it has been applied to various recoupling and decoupling sequences. Hence, it is not too far-fetched to envision that our theoretical framework could be applied to electron paramagnetic resonance pulse sequences employing matching conditions \cite{Jeschke1998a,Rizzato2013}, or to future pulsed MAS-DNP experiments.

\section*{Author contributions}NW, KOT, and GJ designed the research. NW performed all measurements with help from LAV and LS. NW, ABN and NCN developed the theory. All simulation scripts were written by NW, with general help and input from KOT, ABN and LAV. 

\section*{Acknowledgments}
	This article is dedicated to Dr. Anton Ashuiev, stranded in Ukraine at the time of writing. Dr. Daniel Klose and Ren\'e Tschaggelar are acknowledged for building the EPR spectrometer used for this work.  We thank Prof. Matthias Ernst and Prof. Robert G. Griffin for extensive and numerous discussions. Prof. Jan Henrik Ardankj\ae r-Larsen generously provided OX063 for an earlier project. KOT acknowledges the funding support from the French National Research Agency (ANR-20-ERC9-0008) and the  Respore program (petit équipement projet n°339299).	
	NW and GJ acknowledge funding by ETH Zürich grant ETH-48 16-1.

\bibliography{periodic_dnp_lib}

\begin{thebibliography}{51}%
\makeatletter
\providecommand \@ifxundefined [1]{%
 \@ifx{#1\undefined}
}%
\providecommand \@ifnum [1]{%
 \ifnum #1\expandafter \@firstoftwo
 \else \expandafter \@secondoftwo
 \fi
}%
\providecommand \@ifx [1]{%
 \ifx #1\expandafter \@firstoftwo
 \else \expandafter \@secondoftwo
 \fi
}%
\providecommand \natexlab [1]{#1}%
\providecommand \enquote  [1]{``#1''}%
\providecommand \bibnamefont  [1]{#1}%
\providecommand \bibfnamefont [1]{#1}%
\providecommand \citenamefont [1]{#1}%
\providecommand \href@noop [0]{\@secondoftwo}%
\providecommand \href [0]{\begingroup \@sanitize@url \@href}%
\providecommand \@href[1]{\@@startlink{#1}\@@href}%
\providecommand \@@href[1]{\endgroup#1\@@endlink}%
\providecommand \@sanitize@url [0]{\catcode `\\12\catcode `\$12\catcode
  `\&12\catcode `\#12\catcode `\^12\catcode `\_12\catcode `\%12\relax}%
\providecommand \@@startlink[1]{}%
\providecommand \@@endlink[0]{}%
\providecommand \url  [0]{\begingroup\@sanitize@url \@url }%
\providecommand \@url [1]{\endgroup\@href {#1}{\urlprefix }}%
\providecommand \urlprefix  [0]{URL }%
\providecommand \Eprint [0]{\href }%
\providecommand \doibase [0]{http://dx.doi.org/}%
\providecommand \selectlanguage [0]{\@gobble}%
\providecommand \bibinfo  [0]{\@secondoftwo}%
\providecommand \bibfield  [0]{\@secondoftwo}%
\providecommand \translation [1]{[#1]}%
\providecommand \BibitemOpen [0]{}%
\providecommand \bibitemStop [0]{}%
\providecommand \bibitemNoStop [0]{.\EOS\space}%
\providecommand \EOS [0]{\spacefactor3000\relax}%
\providecommand \BibitemShut  [1]{\csname bibitem#1\endcsname}%
\let\auto@bib@innerbib\@empty
\bibitem [{\citenamefont {Abragam}\ and\ \citenamefont
  {Goldman}(1978)}]{Abragam1978}%
  \BibitemOpen
  \bibfield  {author} {\bibinfo {author} {\bibfnamefont {A.}~\bibnamefont
  {Abragam}}\ and\ \bibinfo {author} {\bibfnamefont {M.}~\bibnamefont
  {Goldman}},\ }\bibfield  {title} {\enquote {\bibinfo {title} {{Principles of
  dynamic nuclear polarisation}},}\ }\href {\doibase
  10.1088/0034-4885/41/3/002} {\bibfield  {journal} {\bibinfo  {journal}
  {Reports Prog. Phys.}\ }\textbf {\bibinfo {volume} {41}},\ \bibinfo {pages}
  {395--467} (\bibinfo {year} {1978})}\BibitemShut {NoStop}%
\bibitem [{\citenamefont {{Lilly Thankamony}}\ \emph
  {et~al.}(2017)\citenamefont {{Lilly Thankamony}}, \citenamefont {Wittmann},
  \citenamefont {Kaushik},\ and\ \citenamefont
  {Corzilius}}]{LillyThankamony2017}%
  \BibitemOpen
  \bibfield  {author} {\bibinfo {author} {\bibfnamefont {A.~S.}\ \bibnamefont
  {{Lilly Thankamony}}}, \bibinfo {author} {\bibfnamefont {J.~J.}\ \bibnamefont
  {Wittmann}}, \bibinfo {author} {\bibfnamefont {M.}~\bibnamefont {Kaushik}}, \
  and\ \bibinfo {author} {\bibfnamefont {B.}~\bibnamefont {Corzilius}},\
  }\bibfield  {title} {\enquote {\bibinfo {title} {{Dynamic nuclear
  polarization for sensitivity enhancement in modern solid-state NMR}},}\
  }\href {\doibase 10.1016/j.pnmrs.2017.06.002} {\bibfield  {journal} {\bibinfo
   {journal} {Prog. Nucl. Magn. Reson. Spectrosc.}\ }\textbf {\bibinfo {volume}
  {102-103}},\ \bibinfo {pages} {120--195} (\bibinfo {year}
  {2017})}\BibitemShut {NoStop}%
\bibitem [{\citenamefont {Overhauser}(1953)}]{Overhauser1953}%
  \BibitemOpen
  \bibfield  {author} {\bibinfo {author} {\bibfnamefont {A.~W.}\ \bibnamefont
  {Overhauser}},\ }\bibfield  {title} {\enquote {\bibinfo {title}
  {{Polarization of nuclei in metals}},}\ }\href {\doibase
  10.1103/PhysRev.92.411} {\bibfield  {journal} {\bibinfo  {journal} {Phys.
  Rev.}\ }\textbf {\bibinfo {volume} {92}},\ \bibinfo {pages} {411--415}
  (\bibinfo {year} {1953})}\BibitemShut {NoStop}%
\bibitem [{\citenamefont {Ardenkjaer-Larsen}\ \emph {et~al.}(2003)\citenamefont
  {Ardenkjaer-Larsen}, \citenamefont {Fridlund}, \citenamefont {Gram},
  \citenamefont {Hansson}, \citenamefont {Hansson}, \citenamefont {Lerche},
  \citenamefont {Servin}, \citenamefont {Thaning},\ and\ \citenamefont
  {Golman}}]{Ardenkjaer-Larsen2003}%
  \BibitemOpen
  \bibfield  {author} {\bibinfo {author} {\bibfnamefont {J.~H.}\ \bibnamefont
  {Ardenkjaer-Larsen}}, \bibinfo {author} {\bibfnamefont {B.}~\bibnamefont
  {Fridlund}}, \bibinfo {author} {\bibfnamefont {A.}~\bibnamefont {Gram}},
  \bibinfo {author} {\bibfnamefont {G.}~\bibnamefont {Hansson}}, \bibinfo
  {author} {\bibfnamefont {L.}~\bibnamefont {Hansson}}, \bibinfo {author}
  {\bibfnamefont {M.~H.}\ \bibnamefont {Lerche}}, \bibinfo {author}
  {\bibfnamefont {R.}~\bibnamefont {Servin}}, \bibinfo {author} {\bibfnamefont
  {M.}~\bibnamefont {Thaning}}, \ and\ \bibinfo {author} {\bibfnamefont
  {K.}~\bibnamefont {Golman}},\ }\bibfield  {title} {\enquote {\bibinfo {title}
  {{Increase in signal-to-noise ratio of {\textgreater} 10,000 times in
  liquid-state NMR}},}\ }\href {\doibase 10.1073/pnas.1733835100} {\bibfield
  {journal} {\bibinfo  {journal} {Proc. Natl. Acad. Sci.}\ }\textbf {\bibinfo
  {volume} {100}},\ \bibinfo {pages} {10158--10163} (\bibinfo {year}
  {2003})}\BibitemShut {NoStop}%
\bibitem [{\citenamefont {Becerra}\ \emph {et~al.}(1993)\citenamefont
  {Becerra}, \citenamefont {Gerfen}, \citenamefont {Temkin}, \citenamefont
  {Singel},\ and\ \citenamefont {Griffin}}]{Becerra1993}%
  \BibitemOpen
  \bibfield  {author} {\bibinfo {author} {\bibfnamefont {L.~R.}\ \bibnamefont
  {Becerra}}, \bibinfo {author} {\bibfnamefont {G.~J.}\ \bibnamefont {Gerfen}},
  \bibinfo {author} {\bibfnamefont {R.~J.}\ \bibnamefont {Temkin}}, \bibinfo
  {author} {\bibfnamefont {D.~J.}\ \bibnamefont {Singel}}, \ and\ \bibinfo
  {author} {\bibfnamefont {R.~G.}\ \bibnamefont {Griffin}},\ }\bibfield
  {title} {\enquote {\bibinfo {title} {{Dynamic nuclear polarization with a
  cyclotron resonance maser at 5 T}},}\ }\href {\doibase
  10.1103/PhysRevLett.71.3561} {\bibfield  {journal} {\bibinfo  {journal}
  {Phys. Rev. Lett.}\ }\textbf {\bibinfo {volume} {71}},\ \bibinfo {pages}
  {3561--3564} (\bibinfo {year} {1993})}\BibitemShut {NoStop}%
\bibitem [{\citenamefont {Ardenkj{\ae}r-Larsen}\ \emph
  {et~al.}(2018)\citenamefont {Ardenkj{\ae}r-Larsen}, \citenamefont {Bowen},
  \citenamefont {Petersen}, \citenamefont {Rybalko}, \citenamefont {Vinding},
  \citenamefont {Ullisch},\ and\ \citenamefont {Nielsen}}]{ArdenkjrLarsen2018}%
  \BibitemOpen
  \bibfield  {author} {\bibinfo {author} {\bibfnamefont {J.~H.}\ \bibnamefont
  {Ardenkj{\ae}r-Larsen}}, \bibinfo {author} {\bibfnamefont {S.}~\bibnamefont
  {Bowen}}, \bibinfo {author} {\bibfnamefont {J.~R.}\ \bibnamefont {Petersen}},
  \bibinfo {author} {\bibfnamefont {O.}~\bibnamefont {Rybalko}}, \bibinfo
  {author} {\bibfnamefont {M.~S.}\ \bibnamefont {Vinding}}, \bibinfo {author}
  {\bibfnamefont {M.}~\bibnamefont {Ullisch}}, \ and\ \bibinfo {author}
  {\bibfnamefont {N.~C.}\ \bibnamefont {Nielsen}},\ }\bibfield  {title}
  {\enquote {\bibinfo {title} {Cryogen-free dissolution dynamic nuclear
  polarization polarizer operating at 3.35 t, 6.70 t, and 10.1 t},}\ }\href
  {\doibase 10.1002/mrm.27537} {\bibfield  {journal} {\bibinfo  {journal}
  {Magnetic Resonance in Medicine}\ }\textbf {\bibinfo {volume} {81}},\
  \bibinfo {pages} {2184--2194} (\bibinfo {year} {2018})}\BibitemShut {NoStop}%
\bibitem [{\citenamefont {Jannin}\ \emph {et~al.}(2019)\citenamefont {Jannin},
  \citenamefont {Dumez}, \citenamefont {Giraudeau},\ and\ \citenamefont
  {Kurzbach}}]{Jannin2019a}%
  \BibitemOpen
  \bibfield  {author} {\bibinfo {author} {\bibfnamefont {S.}~\bibnamefont
  {Jannin}}, \bibinfo {author} {\bibfnamefont {J.~N.}\ \bibnamefont {Dumez}},
  \bibinfo {author} {\bibfnamefont {P.}~\bibnamefont {Giraudeau}}, \ and\
  \bibinfo {author} {\bibfnamefont {D.}~\bibnamefont {Kurzbach}},\ }\bibfield
  {title} {\enquote {\bibinfo {title} {{Application and methodology of
  dissolution dynamic nuclear polarization in physical, chemical and biological
  contexts}},}\ }\href {\doibase 10.1016/j.jmr.2019.06.001} {\bibfield
  {journal} {\bibinfo  {journal} {J. Magn. Reson.}\ }\textbf {\bibinfo {volume}
  {305}},\ \bibinfo {pages} {41--50} (\bibinfo {year} {2019})}\BibitemShut
  {NoStop}%
\bibitem [{\citenamefont {Tan}\ \emph {et~al.}(2019{\natexlab{a}})\citenamefont
  {Tan}, \citenamefont {Jawla}, \citenamefont {Temkin},\ and\ \citenamefont
  {Griffin}}]{Tan2019a}%
  \BibitemOpen
  \bibfield  {author} {\bibinfo {author} {\bibfnamefont {K.~O.}\ \bibnamefont
  {Tan}}, \bibinfo {author} {\bibfnamefont {S.}~\bibnamefont {Jawla}}, \bibinfo
  {author} {\bibfnamefont {R.~J.}\ \bibnamefont {Temkin}}, \ and\ \bibinfo
  {author} {\bibfnamefont {R.~G.}\ \bibnamefont {Griffin}},\ }\bibfield
  {title} {\enquote {\bibinfo {title} {{Pulsed dynamic nuclear
  polarization}},}\ }\href {\doibase 10.1002/9780470034590.emrstm1551}
  {\bibfield  {journal} {\bibinfo  {journal} {eMagRes}\ }\textbf {\bibinfo
  {volume} {8}},\ \bibinfo {pages} {339--352} (\bibinfo {year}
  {2019}{\natexlab{a}})}\BibitemShut {NoStop}%
\bibitem [{\citenamefont {Berruyer}\ \emph {et~al.}(2020)\citenamefont
  {Berruyer}, \citenamefont {Bj\"{o}rgvinsd{\'{o}}ttir}, \citenamefont
  {Bertarello}, \citenamefont {Stevanato}, \citenamefont {Rao}, \citenamefont
  {Karthikeyan}, \citenamefont {Casano}, \citenamefont {Ouari}, \citenamefont
  {Lelli}, \citenamefont {Reiter}, \citenamefont {Engelke},\ and\ \citenamefont
  {Emsley}}]{Berruyer2020}%
  \BibitemOpen
  \bibfield  {author} {\bibinfo {author} {\bibfnamefont {P.}~\bibnamefont
  {Berruyer}}, \bibinfo {author} {\bibfnamefont {S.}~\bibnamefont
  {Bj\"{o}rgvinsd{\'{o}}ttir}}, \bibinfo {author} {\bibfnamefont
  {A.}~\bibnamefont {Bertarello}}, \bibinfo {author} {\bibfnamefont
  {G.}~\bibnamefont {Stevanato}}, \bibinfo {author} {\bibfnamefont
  {Y.}~\bibnamefont {Rao}}, \bibinfo {author} {\bibfnamefont {G.}~\bibnamefont
  {Karthikeyan}}, \bibinfo {author} {\bibfnamefont {G.}~\bibnamefont {Casano}},
  \bibinfo {author} {\bibfnamefont {O.}~\bibnamefont {Ouari}}, \bibinfo
  {author} {\bibfnamefont {M.}~\bibnamefont {Lelli}}, \bibinfo {author}
  {\bibfnamefont {C.}~\bibnamefont {Reiter}}, \bibinfo {author} {\bibfnamefont
  {F.}~\bibnamefont {Engelke}}, \ and\ \bibinfo {author} {\bibfnamefont
  {L.}~\bibnamefont {Emsley}},\ }\bibfield  {title} {\enquote {\bibinfo {title}
  {Dynamic nuclear polarization enhancement of 200 at 21.15 t enabled by 65
  {kHz} magic angle spinning},}\ }\href {\doibase 10.1021/acs.jpclett.0c02493}
  {\bibfield  {journal} {\bibinfo  {journal} {The Journal of Physical Chemistry
  Letters}\ }\textbf {\bibinfo {volume} {11}},\ \bibinfo {pages} {8386--8391}
  (\bibinfo {year} {2020})}\BibitemShut {NoStop}%
\bibitem [{\citenamefont {Cai}\ \emph {et~al.}(2021)\citenamefont {Cai},
  \citenamefont {Paioni}, \citenamefont {Adler}, \citenamefont {Yao},
  \citenamefont {Zhang}, \citenamefont {Beriashvili}, \citenamefont {Safeer},
  \citenamefont {Gurinov}, \citenamefont {Rockenbauer}, \citenamefont {Song},
  \citenamefont {Baldus},\ and\ \citenamefont {Liu}}]{Cai2021}%
  \BibitemOpen
  \bibfield  {author} {\bibinfo {author} {\bibfnamefont {X.}~\bibnamefont
  {Cai}}, \bibinfo {author} {\bibfnamefont {A.~L.}\ \bibnamefont {Paioni}},
  \bibinfo {author} {\bibfnamefont {A.}~\bibnamefont {Adler}}, \bibinfo
  {author} {\bibfnamefont {R.}~\bibnamefont {Yao}}, \bibinfo {author}
  {\bibfnamefont {W.}~\bibnamefont {Zhang}}, \bibinfo {author} {\bibfnamefont
  {D.}~\bibnamefont {Beriashvili}}, \bibinfo {author} {\bibfnamefont
  {A.}~\bibnamefont {Safeer}}, \bibinfo {author} {\bibfnamefont
  {A.}~\bibnamefont {Gurinov}}, \bibinfo {author} {\bibfnamefont
  {A.}~\bibnamefont {Rockenbauer}}, \bibinfo {author} {\bibfnamefont
  {Y.}~\bibnamefont {Song}}, \bibinfo {author} {\bibfnamefont {M.}~\bibnamefont
  {Baldus}}, \ and\ \bibinfo {author} {\bibfnamefont {Y.}~\bibnamefont {Liu}},\
  }\bibfield  {title} {\enquote {\bibinfo {title} {Highly efficient
  trityl-nitroxide biradicals for biomolecular high-field dynamic nuclear
  polarization},}\ }\href {\doibase 10.1002/chem.202102253} {\bibfield
  {journal} {\bibinfo  {journal} {Chemistry {\textendash} A European Journal}\
  }\textbf {\bibinfo {volume} {27}},\ \bibinfo {pages} {12758--12762} (\bibinfo
  {year} {2021})}\BibitemShut {NoStop}%
\bibitem [{\citenamefont {Abragam}\ and\ \citenamefont
  {Proctor}(1958)}]{Abragam1958}%
  \BibitemOpen
  \bibfield  {author} {\bibinfo {author} {\bibfnamefont {A.}~\bibnamefont
  {Abragam}}\ and\ \bibinfo {author} {\bibfnamefont {W.~G.}\ \bibnamefont
  {Proctor}},\ }\bibfield  {title} {\enquote {\bibinfo {title} {{Une nouvelle
  methode de polarisation dynamique des noyaux atomiques dans les solides.}}}\
  }\href@noop {} {\bibfield  {journal} {\bibinfo  {journal} {Comp. Rend. Acad.
  Sci.}\ }\textbf {\bibinfo {volume} {246}},\ \bibinfo {pages} {2253--2256}
  (\bibinfo {year} {1958})}\BibitemShut {NoStop}%
\bibitem [{\citenamefont {C.D.Jeffries}(1957)}]{C.D.Jeffries1957}%
  \BibitemOpen
  \bibfield  {author} {\bibinfo {author} {\bibnamefont {C.D.Jeffries}},\
  }\bibfield  {title} {\enquote {\bibinfo {title} {{Polarisation of Nuclei by
  Resonance Saturation in Paramagnetic Crystals}},}\ }\href@noop {} {\bibfield
  {journal} {\bibinfo  {journal} {Phys. Rev.}\ }\textbf {\bibinfo {volume}
  {106}},\ \bibinfo {pages} {164--165} (\bibinfo {year} {1957})}\BibitemShut
  {NoStop}%
\bibitem [{\citenamefont {Hwang}\ and\ \citenamefont {Hill}(1967)}]{Hwang1967}%
  \BibitemOpen
  \bibfield  {author} {\bibinfo {author} {\bibfnamefont {C.~F.}\ \bibnamefont
  {Hwang}}\ and\ \bibinfo {author} {\bibfnamefont {D.~A.}\ \bibnamefont
  {Hill}},\ }\bibfield  {title} {\enquote {\bibinfo {title} {{New effect in
  dynamic polarization}},}\ }\href {\doibase 10.1103/PhysRevLett.18.110}
  {\bibfield  {journal} {\bibinfo  {journal} {Phys. Rev. Lett.}\ }\textbf
  {\bibinfo {volume} {18}},\ \bibinfo {pages} {110--112} (\bibinfo {year}
  {1967})}\BibitemShut {NoStop}%
\bibitem [{\citenamefont {Kessenikh}\ \emph {et~al.}(1963)\citenamefont
  {Kessenikh}, \citenamefont {Luschikov}, \citenamefont {Manekov},\ and\
  \citenamefont {Taran}}]{Kessenikh1963}%
  \BibitemOpen
  \bibfield  {author} {\bibinfo {author} {\bibfnamefont {A.}~\bibnamefont
  {Kessenikh}}, \bibinfo {author} {\bibfnamefont {V.}~\bibnamefont
  {Luschikov}}, \bibinfo {author} {\bibfnamefont {A.}~\bibnamefont {Manekov}},
  \ and\ \bibinfo {author} {\bibfnamefont {Y.~V.}\ \bibnamefont {Taran}},\
  }\bibfield  {title} {\enquote {\bibinfo {title} {Proton polarization in
  irradiated polyethylenes},}\ }\href@noop {} {\bibfield  {journal} {\bibinfo
  {journal} {Sov. Physics–Solid State}\ }\textbf {\bibinfo {volume} {5}},\
  \bibinfo {pages} {321--329} (\bibinfo {year} {1963})}\BibitemShut {NoStop}%
\bibitem [{\citenamefont {Provotorov}(1962)}]{Provotorov1962}%
  \BibitemOpen
  \bibfield  {author} {\bibinfo {author} {\bibfnamefont {B.~N.}\ \bibnamefont
  {Provotorov}},\ }\bibfield  {title} {\enquote {\bibinfo {title} {{Magnetic
  Resonance Saturation in Crystals}},}\ }\href@noop {} {\bibfield  {journal}
  {\bibinfo  {journal} {Sov. Phys. JETP}\ }\textbf {\bibinfo {volume} {14}},\
  \bibinfo {pages} {1126--1131} (\bibinfo {year} {1962})}\BibitemShut {NoStop}%
\bibitem [{\citenamefont {Borghini}(1968)}]{Borghini1968}%
  \BibitemOpen
  \bibfield  {author} {\bibinfo {author} {\bibfnamefont {M.}~\bibnamefont
  {Borghini}},\ }\bibfield  {title} {\enquote {\bibinfo {title}
  {{Spin-temperature model of nuclear dynamic polarization using free
  radicals}},}\ }\href {\doibase 10.1103/PhysRevLett.20.419} {\bibfield
  {journal} {\bibinfo  {journal} {Phys. Rev. Lett.}\ }\textbf {\bibinfo
  {volume} {20}},\ \bibinfo {pages} {419--421} (\bibinfo {year}
  {1968})}\BibitemShut {NoStop}%
\bibitem [{\citenamefont {Soetbeer}\ \emph {et~al.}(2018)\citenamefont
  {Soetbeer}, \citenamefont {Gast}, \citenamefont {Walish}, \citenamefont
  {Zhao}, \citenamefont {George}, \citenamefont {Yang}, \citenamefont {Swager},
  \citenamefont {Griffin},\ and\ \citenamefont {Mathies}}]{Soetbeer2018}%
  \BibitemOpen
  \bibfield  {author} {\bibinfo {author} {\bibfnamefont {J.}~\bibnamefont
  {Soetbeer}}, \bibinfo {author} {\bibfnamefont {P.}~\bibnamefont {Gast}},
  \bibinfo {author} {\bibfnamefont {J.~J.}\ \bibnamefont {Walish}}, \bibinfo
  {author} {\bibfnamefont {Y.}~\bibnamefont {Zhao}}, \bibinfo {author}
  {\bibfnamefont {C.}~\bibnamefont {George}}, \bibinfo {author} {\bibfnamefont
  {C.}~\bibnamefont {Yang}}, \bibinfo {author} {\bibfnamefont {T.~M.}\
  \bibnamefont {Swager}}, \bibinfo {author} {\bibfnamefont {R.~G.}\
  \bibnamefont {Griffin}}, \ and\ \bibinfo {author} {\bibfnamefont
  {G.}~\bibnamefont {Mathies}},\ }\bibfield  {title} {\enquote {\bibinfo
  {title} {Conformation of bis-nitroxide polarizing agents by multi-frequency
  {EPR} spectroscopy},}\ }\href {\doibase 10.1039/c8cp05236k} {\bibfield
  {journal} {\bibinfo  {journal} {Physical Chemistry Chemical Physics}\
  }\textbf {\bibinfo {volume} {20}},\ \bibinfo {pages} {25506--25517} (\bibinfo
  {year} {2018})}\BibitemShut {NoStop}%
\bibitem [{\citenamefont {Henstra}\ \emph {et~al.}(1988)\citenamefont
  {Henstra}, \citenamefont {Dirksen}, \citenamefont {Schmidt},\ and\
  \citenamefont {Wenckebach}}]{Henstra1988}%
  \BibitemOpen
  \bibfield  {author} {\bibinfo {author} {\bibfnamefont {A.}~\bibnamefont
  {Henstra}}, \bibinfo {author} {\bibfnamefont {P.}~\bibnamefont {Dirksen}},
  \bibinfo {author} {\bibfnamefont {J.}~\bibnamefont {Schmidt}}, \ and\
  \bibinfo {author} {\bibfnamefont {W.~T.}\ \bibnamefont {Wenckebach}},\
  }\bibfield  {title} {\enquote {\bibinfo {title} {{Nuclear spin orientation
  via electron spin locking (NOVEL)}},}\ }\href {\doibase
  10.1016/0022-2364(88)90190-4} {\bibfield  {journal} {\bibinfo  {journal}
  {Journal of Magnetic Resonance (1969)}\ }\textbf {\bibinfo {volume} {77}},\
  \bibinfo {pages} {389--393} (\bibinfo {year} {1988})}\BibitemShut {NoStop}%
\bibitem [{\citenamefont {Can}\ \emph {et~al.}(2015)\citenamefont {Can},
  \citenamefont {Walish}, \citenamefont {Swager},\ and\ \citenamefont
  {Griffin}}]{Can2015a}%
  \BibitemOpen
  \bibfield  {author} {\bibinfo {author} {\bibfnamefont {T.~V.}\ \bibnamefont
  {Can}}, \bibinfo {author} {\bibfnamefont {J.~J.}\ \bibnamefont {Walish}},
  \bibinfo {author} {\bibfnamefont {T.~M.}\ \bibnamefont {Swager}}, \ and\
  \bibinfo {author} {\bibfnamefont {R.~G.}\ \bibnamefont {Griffin}},\
  }\bibfield  {title} {\enquote {\bibinfo {title} {{Time domain DNP with the
  NOVEL sequence}},}\ }\href {\doibase 10.1063/1.4927087} {\bibfield  {journal}
  {\bibinfo  {journal} {J. Chem. Phys.}\ }\textbf {\bibinfo {volume} {143}},\
  \bibinfo {pages} {054201} (\bibinfo {year} {2015})}\BibitemShut {NoStop}%
\bibitem [{\citenamefont {Mathies}\ \emph {et~al.}(2016)\citenamefont
  {Mathies}, \citenamefont {Jain}, \citenamefont {Reese},\ and\ \citenamefont
  {Griffin}}]{Mathies2016}%
  \BibitemOpen
  \bibfield  {author} {\bibinfo {author} {\bibfnamefont {G.}~\bibnamefont
  {Mathies}}, \bibinfo {author} {\bibfnamefont {S.}~\bibnamefont {Jain}},
  \bibinfo {author} {\bibfnamefont {M.}~\bibnamefont {Reese}}, \ and\ \bibinfo
  {author} {\bibfnamefont {R.~G.}\ \bibnamefont {Griffin}},\ }\bibfield
  {title} {\enquote {\bibinfo {title} {{Pulsed Dynamic Nuclear Polarization
  with Trityl Radicals}},}\ }\href {\doibase 10.1021/acs.jpclett.5b02720}
  {\bibfield  {journal} {\bibinfo  {journal} {J. Phys. Chem. Lett.}\ }\textbf
  {\bibinfo {volume} {7}},\ \bibinfo {pages} {111--116} (\bibinfo {year}
  {2016})}\BibitemShut {NoStop}%
\bibitem [{\citenamefont {Can}\ \emph {et~al.}(2017{\natexlab{a}})\citenamefont
  {Can}, \citenamefont {Weber}, \citenamefont {Walish}, \citenamefont
  {Swager},\ and\ \citenamefont {Griffin}}]{Can2017b}%
  \BibitemOpen
  \bibfield  {author} {\bibinfo {author} {\bibfnamefont {T.~V.}\ \bibnamefont
  {Can}}, \bibinfo {author} {\bibfnamefont {R.~T.}\ \bibnamefont {Weber}},
  \bibinfo {author} {\bibfnamefont {J.~J.}\ \bibnamefont {Walish}}, \bibinfo
  {author} {\bibfnamefont {T.~M.}\ \bibnamefont {Swager}}, \ and\ \bibinfo
  {author} {\bibfnamefont {R.~G.}\ \bibnamefont {Griffin}},\ }\bibfield
  {title} {\enquote {\bibinfo {title} {{Ramped-amplitude NOVEL}},}\ }\href
  {\doibase 10.1063/1.4980155} {\bibfield  {journal} {\bibinfo  {journal} {J.
  Chem. Phys.}\ }\textbf {\bibinfo {volume} {146}},\ \bibinfo {pages} {154204}
  (\bibinfo {year} {2017}{\natexlab{a}})}\BibitemShut {NoStop}%
\bibitem [{\citenamefont {Jain}, \citenamefont {Mathies},\ and\ \citenamefont
  {Griffin}(2017)}]{Jain2017}%
  \BibitemOpen
  \bibfield  {author} {\bibinfo {author} {\bibfnamefont {S.~K.}\ \bibnamefont
  {Jain}}, \bibinfo {author} {\bibfnamefont {G.}~\bibnamefont {Mathies}}, \
  and\ \bibinfo {author} {\bibfnamefont {R.~G.}\ \bibnamefont {Griffin}},\
  }\bibfield  {title} {\enquote {\bibinfo {title} {{Off-resonance NOVEL}},}\
  }\href {\doibase 10.1063/1.5000528} {\bibfield  {journal} {\bibinfo
  {journal} {J. Chem. Phys.}\ }\textbf {\bibinfo {volume} {147}},\ \bibinfo
  {pages} {164201} (\bibinfo {year} {2017})}\BibitemShut {NoStop}%
\bibitem [{\citenamefont {Henstra}, \citenamefont {Dirksen},\ and\
  \citenamefont {Wenckebach}(1988)}]{Henstra1988a}%
  \BibitemOpen
  \bibfield  {author} {\bibinfo {author} {\bibfnamefont {A.}~\bibnamefont
  {Henstra}}, \bibinfo {author} {\bibfnamefont {P.}~\bibnamefont {Dirksen}}, \
  and\ \bibinfo {author} {\bibfnamefont {W.~T.}\ \bibnamefont {Wenckebach}},\
  }\bibfield  {title} {\enquote {\bibinfo {title} {{Enhanced dynamic nuclear
  polarization by the integrated solid effect}},}\ }\href {\doibase
  10.1016/0375-9601(88)90950-4} {\bibfield  {journal} {\bibinfo  {journal}
  {Phys. Lett. A}\ }\textbf {\bibinfo {volume} {134}},\ \bibinfo {pages}
  {134--136} (\bibinfo {year} {1988})}\BibitemShut {NoStop}%
\bibitem [{\citenamefont {Can}\ \emph {et~al.}(2017{\natexlab{b}})\citenamefont
  {Can}, \citenamefont {Weber}, \citenamefont {Walish}, \citenamefont
  {Swager},\ and\ \citenamefont {Griffin}}]{Can2017}%
  \BibitemOpen
  \bibfield  {author} {\bibinfo {author} {\bibfnamefont {T.~V.}\ \bibnamefont
  {Can}}, \bibinfo {author} {\bibfnamefont {R.~T.}\ \bibnamefont {Weber}},
  \bibinfo {author} {\bibfnamefont {J.~J.}\ \bibnamefont {Walish}}, \bibinfo
  {author} {\bibfnamefont {T.~M.}\ \bibnamefont {Swager}}, \ and\ \bibinfo
  {author} {\bibfnamefont {R.~G.}\ \bibnamefont {Griffin}},\ }\bibfield
  {title} {\enquote {\bibinfo {title} {{Frequency-Swept Integrated Solid
  Effect}},}\ }\href {\doibase 10.1002/anie.201700032} {\bibfield  {journal}
  {\bibinfo  {journal} {Angew. Chemie - Int. Ed.}\ }\textbf {\bibinfo {volume}
  {56}},\ \bibinfo {pages} {6744--6748} (\bibinfo {year}
  {2017}{\natexlab{b}})}\BibitemShut {NoStop}%
\bibitem [{\citenamefont {Tan}\ \emph {et~al.}(2020)\citenamefont {Tan},
  \citenamefont {Weber}, \citenamefont {Can},\ and\ \citenamefont
  {Griffin}}]{Tan2020}%
  \BibitemOpen
  \bibfield  {author} {\bibinfo {author} {\bibfnamefont {K.~O.}\ \bibnamefont
  {Tan}}, \bibinfo {author} {\bibfnamefont {R.~T.}\ \bibnamefont {Weber}},
  \bibinfo {author} {\bibfnamefont {T.~V.}\ \bibnamefont {Can}}, \ and\
  \bibinfo {author} {\bibfnamefont {R.~G.}\ \bibnamefont {Griffin}},\
  }\bibfield  {title} {\enquote {\bibinfo {title} {{Adiabatic Solid Effect}},}\
  }\href {\doibase 10.1021/acs.jpclett.0c00654} {\bibfield  {journal} {\bibinfo
   {journal} {J. Phys. Chem. Lett.}\ }\textbf {\bibinfo {volume} {11}},\
  \bibinfo {pages} {3416--3421} (\bibinfo {year} {2020})}\BibitemShut {NoStop}%
\bibitem [{\citenamefont {Wind}\ \emph {et~al.}(1988)\citenamefont {Wind},
  \citenamefont {Li}, \citenamefont {Lock},\ and\ \citenamefont
  {Maciel}}]{Wind1988}%
  \BibitemOpen
  \bibfield  {author} {\bibinfo {author} {\bibfnamefont {R.~A.}\ \bibnamefont
  {Wind}}, \bibinfo {author} {\bibfnamefont {L.}~\bibnamefont {Li}}, \bibinfo
  {author} {\bibfnamefont {H.}~\bibnamefont {Lock}}, \ and\ \bibinfo {author}
  {\bibfnamefont {G.~E.}\ \bibnamefont {Maciel}},\ }\bibfield  {title}
  {\enquote {\bibinfo {title} {{Dynamic nuclear polarization in the nuclear
  rotating frame}},}\ }\href {\doibase 10.1016/0022-2364(88)90094-7} {\bibfield
   {journal} {\bibinfo  {journal} {J. Magn. Reson.}\ }\textbf {\bibinfo
  {volume} {79}},\ \bibinfo {pages} {577--582} (\bibinfo {year}
  {1988})}\BibitemShut {NoStop}%
\bibitem [{\citenamefont {Weis}\ \emph {et~al.}(2000)\citenamefont {Weis},
  \citenamefont {Bennati}, \citenamefont {Rosay},\ and\ \citenamefont
  {Griffin}}]{Weis2000}%
  \BibitemOpen
  \bibfield  {author} {\bibinfo {author} {\bibfnamefont {V.}~\bibnamefont
  {Weis}}, \bibinfo {author} {\bibfnamefont {M.}~\bibnamefont {Bennati}},
  \bibinfo {author} {\bibfnamefont {M.}~\bibnamefont {Rosay}}, \ and\ \bibinfo
  {author} {\bibfnamefont {R.~G.}\ \bibnamefont {Griffin}},\ }\bibfield
  {title} {\enquote {\bibinfo {title} {{Solid effect in the electron spin
  dressed state: a new approach for dynamic nuclear polarization}},}\ }\href
  {\doibase 10.1063/1.1310599} {\bibfield  {journal} {\bibinfo  {journal} {J.
  Chem. Phys.}\ }\textbf {\bibinfo {volume} {113}},\ \bibinfo {pages}
  {6795--6802} (\bibinfo {year} {2000})}\BibitemShut {NoStop}%
\bibitem [{\citenamefont {Schwartz}\ \emph {et~al.}(2018)\citenamefont
  {Schwartz}, \citenamefont {Scheuer}, \citenamefont {Tratzmiller},
  \citenamefont {Müller}, \citenamefont {Chen}, \citenamefont {Dhand},
  \citenamefont {Wang}, \citenamefont {Müller}, \citenamefont {Naydenov},
  \citenamefont {Jelezko},\ and\ \citenamefont {Plenio}}]{Schwartz2018a}%
  \BibitemOpen
  \bibfield  {author} {\bibinfo {author} {\bibfnamefont {I.}~\bibnamefont
  {Schwartz}}, \bibinfo {author} {\bibfnamefont {J.}~\bibnamefont {Scheuer}},
  \bibinfo {author} {\bibfnamefont {B.}~\bibnamefont {Tratzmiller}}, \bibinfo
  {author} {\bibfnamefont {S.}~\bibnamefont {Müller}}, \bibinfo {author}
  {\bibfnamefont {Q.}~\bibnamefont {Chen}}, \bibinfo {author} {\bibfnamefont
  {I.}~\bibnamefont {Dhand}}, \bibinfo {author} {\bibfnamefont {Z.-Y.}\
  \bibnamefont {Wang}}, \bibinfo {author} {\bibfnamefont {C.}~\bibnamefont
  {Müller}}, \bibinfo {author} {\bibfnamefont {B.}~\bibnamefont {Naydenov}},
  \bibinfo {author} {\bibfnamefont {F.}~\bibnamefont {Jelezko}}, \ and\
  \bibinfo {author} {\bibfnamefont {M.~B.}\ \bibnamefont {Plenio}},\ }\bibfield
   {title} {\enquote {\bibinfo {title} {Robust optical polarization of nuclear
  spin baths using hamiltonian engineering of nitrogen-vacancy center quantum
  dynamics},}\ }\href {\doibase 10.1126/sciadv.aat8978} {\bibfield  {journal}
  {\bibinfo  {journal} {Science Advances}\ }\textbf {\bibinfo {volume} {4}},\
  \bibinfo {pages} {eaat8978} (\bibinfo {year} {2018})}\BibitemShut {NoStop}%
\bibitem [{\citenamefont {Tan}\ \emph {et~al.}(2019{\natexlab{b}})\citenamefont
  {Tan}, \citenamefont {Yang}, \citenamefont {Weber}, \citenamefont {Mathies},\
  and\ \citenamefont {Griffin}}]{OoiTan2019}%
  \BibitemOpen
  \bibfield  {author} {\bibinfo {author} {\bibfnamefont {K.~O.}\ \bibnamefont
  {Tan}}, \bibinfo {author} {\bibfnamefont {C.}~\bibnamefont {Yang}}, \bibinfo
  {author} {\bibfnamefont {R.~T.}\ \bibnamefont {Weber}}, \bibinfo {author}
  {\bibfnamefont {G.}~\bibnamefont {Mathies}}, \ and\ \bibinfo {author}
  {\bibfnamefont {R.~G.}\ \bibnamefont {Griffin}},\ }\bibfield  {title}
  {\enquote {\bibinfo {title} {Time-optimized pulsed dynamic nuclear
  polarization},}\ }\href {\doibase 10.1126/sciadv.aav6909} {\bibfield
  {journal} {\bibinfo  {journal} {Science Advances}\ }\textbf {\bibinfo
  {volume} {5}},\ \bibinfo {pages} {eaav6909} (\bibinfo {year}
  {2019}{\natexlab{b}})}\BibitemShut {NoStop}%
\bibitem [{\citenamefont {Redrouthu}\ and\ \citenamefont
  {Mathies}(2022)}]{Mathies2022}%
  \BibitemOpen
  \bibfield  {author} {\bibinfo {author} {\bibfnamefont {V.~S.}\ \bibnamefont
  {Redrouthu}}\ and\ \bibinfo {author} {\bibfnamefont {G.}~\bibnamefont
  {Mathies}},\ }\bibfield  {title} {\enquote {\bibinfo {title} {Efficient
  pulsed dynamic nuclear polarization with the x-inverse-x sequence},}\ }\href
  {\doibase 10.1021/jacs.1c09900} {\bibfield  {journal} {\bibinfo  {journal}
  {Journal of the American Chemical Society}\ }\textbf {\bibinfo {volume}
  {144}},\ \bibinfo {pages} {1513--1516} (\bibinfo {year} {2022})}\BibitemShut
  {NoStop}%
\bibitem [{\citenamefont {Shankar}\ \emph {et~al.}(2017)\citenamefont
  {Shankar}, \citenamefont {Ernst}, \citenamefont {Madhu}, \citenamefont
  {Vosegaard}, \citenamefont {Nielsen},\ and\ \citenamefont
  {Nielsen}}]{Shankar2017}%
  \BibitemOpen
  \bibfield  {author} {\bibinfo {author} {\bibfnamefont {R.}~\bibnamefont
  {Shankar}}, \bibinfo {author} {\bibfnamefont {M.}~\bibnamefont {Ernst}},
  \bibinfo {author} {\bibfnamefont {P.~K.}\ \bibnamefont {Madhu}}, \bibinfo
  {author} {\bibfnamefont {T.}~\bibnamefont {Vosegaard}}, \bibinfo {author}
  {\bibfnamefont {N.~C.}\ \bibnamefont {Nielsen}}, \ and\ \bibinfo {author}
  {\bibfnamefont {A.~B.}\ \bibnamefont {Nielsen}},\ }\bibfield  {title}
  {\enquote {\bibinfo {title} {A general theoretical description of the
  influence of isotropic chemical shift in dipolar recoupling experiments for
  solid-state {NMR}},}\ }\href {\doibase 10.1063/1.4979123} {\bibfield
  {journal} {\bibinfo  {journal} {The Journal of Chemical Physics}\ }\textbf
  {\bibinfo {volume} {146}},\ \bibinfo {pages} {134105} (\bibinfo {year}
  {2017})}\BibitemShut {NoStop}%
\bibitem [{\citenamefont {Nielsen}\ \emph {et~al.}(2019)\citenamefont
  {Nielsen}, \citenamefont {Hansen}, \citenamefont {Andersen},\ and\
  \citenamefont {Vosegaard}}]{Nielsen2019}%
  \BibitemOpen
  \bibfield  {author} {\bibinfo {author} {\bibfnamefont {A.~B.}\ \bibnamefont
  {Nielsen}}, \bibinfo {author} {\bibfnamefont {M.~R.}\ \bibnamefont {Hansen}},
  \bibinfo {author} {\bibfnamefont {J.~E.}\ \bibnamefont {Andersen}}, \ and\
  \bibinfo {author} {\bibfnamefont {T.}~\bibnamefont {Vosegaard}},\ }\bibfield
  {title} {\enquote {\bibinfo {title} {Single-spin vector analysis of strongly
  coupled nuclei in {TOCSY} {NMR} experiments},}\ }\href {\doibase
  10.1063/1.5123046} {\bibfield  {journal} {\bibinfo  {journal} {The Journal of
  Chemical Physics}\ }\textbf {\bibinfo {volume} {151}},\ \bibinfo {pages}
  {134117} (\bibinfo {year} {2019})}\BibitemShut {NoStop}%
\bibitem [{\citenamefont {Wolfe}(1973)}]{Wolfe1973}%
  \BibitemOpen
  \bibfield  {author} {\bibinfo {author} {\bibfnamefont {J.~P.}\ \bibnamefont
  {Wolfe}},\ }\bibfield  {title} {\enquote {\bibinfo {title} {{Direct
  Observation of a Nuclear Spin Diffusion Barrier}},}\ }\href {\doibase
  10.1103/PhysRevLett.31.907} {\bibfield  {journal} {\bibinfo  {journal} {Phys.
  Rev. Lett.}\ }\textbf {\bibinfo {volume} {31}},\ \bibinfo {pages} {907--910}
  (\bibinfo {year} {1973})}\BibitemShut {NoStop}%
\bibitem [{\citenamefont {Tan}\ \emph {et~al.}(2019{\natexlab{c}})\citenamefont
  {Tan}, \citenamefont {Mardini}, \citenamefont {Yang}, \citenamefont
  {Ardenkj{\ae}r-Larsen},\ and\ \citenamefont {Griffin}}]{Tan2019}%
  \BibitemOpen
  \bibfield  {author} {\bibinfo {author} {\bibfnamefont {K.~O.}\ \bibnamefont
  {Tan}}, \bibinfo {author} {\bibfnamefont {M.}~\bibnamefont {Mardini}},
  \bibinfo {author} {\bibfnamefont {C.}~\bibnamefont {Yang}}, \bibinfo {author}
  {\bibfnamefont {J.~H.}\ \bibnamefont {Ardenkj{\ae}r-Larsen}}, \ and\ \bibinfo
  {author} {\bibfnamefont {R.~G.}\ \bibnamefont {Griffin}},\ }\bibfield
  {title} {\enquote {\bibinfo {title} {{Three-spin solid effect and the spin
  diffusion barrier in amorphous solids}},}\ }\href {\doibase
  10.1126/sciadv.aax2743} {\bibfield  {journal} {\bibinfo  {journal} {Sci.
  Adv.}\ }\textbf {\bibinfo {volume} {5}},\ \bibinfo {pages} {eaax2743}
  (\bibinfo {year} {2019}{\natexlab{c}})}\BibitemShut {NoStop}%
\bibitem [{\citenamefont {Jain}\ \emph {et~al.}(2021)\citenamefont {Jain},
  \citenamefont {Yu}, \citenamefont {Wilson}, \citenamefont {Tabassum},
  \citenamefont {Freedman},\ and\ \citenamefont {Han}}]{Jain2021}%
  \BibitemOpen
  \bibfield  {author} {\bibinfo {author} {\bibfnamefont {S.~K.}\ \bibnamefont
  {Jain}}, \bibinfo {author} {\bibfnamefont {C.~J.}\ \bibnamefont {Yu}},
  \bibinfo {author} {\bibfnamefont {C.~B.}\ \bibnamefont {Wilson}}, \bibinfo
  {author} {\bibfnamefont {T.}~\bibnamefont {Tabassum}}, \bibinfo {author}
  {\bibfnamefont {D.~E.}\ \bibnamefont {Freedman}}, \ and\ \bibinfo {author}
  {\bibfnamefont {S.}~\bibnamefont {Han}},\ }\bibfield  {title} {\enquote
  {\bibinfo {title} {{Dynamic Nuclear Polarization with Vanadium(IV) Metal
  Centers}},}\ }\href {\doibase 10.1016/j.chempr.2020.10.021} {\bibfield
  {journal} {\bibinfo  {journal} {Chem}\ }\textbf {\bibinfo {volume} {7}},\
  \bibinfo {pages} {421--435} (\bibinfo {year} {2021})}\BibitemShut {NoStop}%
\bibitem [{\citenamefont {Stern}\ \emph {et~al.}(2021)\citenamefont {Stern},
  \citenamefont {Cousin}, \citenamefont {Mentink-Vigier}, \citenamefont
  {Pinon}, \citenamefont {Elliott}, \citenamefont {Cala},\ and\ \citenamefont
  {Jannin}}]{Stern2021}%
  \BibitemOpen
  \bibfield  {author} {\bibinfo {author} {\bibfnamefont {Q.}~\bibnamefont
  {Stern}}, \bibinfo {author} {\bibfnamefont {S.~F.}\ \bibnamefont {Cousin}},
  \bibinfo {author} {\bibfnamefont {F.}~\bibnamefont {Mentink-Vigier}},
  \bibinfo {author} {\bibfnamefont {A.~C.}\ \bibnamefont {Pinon}}, \bibinfo
  {author} {\bibfnamefont {S.~J.}\ \bibnamefont {Elliott}}, \bibinfo {author}
  {\bibfnamefont {O.}~\bibnamefont {Cala}}, \ and\ \bibinfo {author}
  {\bibfnamefont {S.}~\bibnamefont {Jannin}},\ }\bibfield  {title} {\enquote
  {\bibinfo {title} {{Direct observation of hyperpolarization breaking through
  the spin diffusion barrier}},}\ }\href {\doibase 10.1126/sciadv.abf5735}
  {\bibfield  {journal} {\bibinfo  {journal} {Sci. Adv.}\ }\textbf {\bibinfo
  {volume} {7}},\ \bibinfo {pages} {1--14} (\bibinfo {year}
  {2021})}\BibitemShut {NoStop}%
\bibitem [{\citenamefont {Bl{\"{u}}mich}\ and\ \citenamefont
  {Spiess}(1985)}]{Blumich1985}%
  \BibitemOpen
  \bibfield  {author} {\bibinfo {author} {\bibfnamefont {B.}~\bibnamefont
  {Bl{\"{u}}mich}}\ and\ \bibinfo {author} {\bibfnamefont {H.~W.}\ \bibnamefont
  {Spiess}},\ }\bibfield  {title} {\enquote {\bibinfo {title} {{Quaternions as
  a practical tool for the evaluation of composite rotations}},}\ }\href
  {\doibase 10.1016/0022-2364(85)90091-5} {\bibfield  {journal} {\bibinfo
  {journal} {J. Magn. Reson.}\ }\textbf {\bibinfo {volume} {61}},\ \bibinfo
  {pages} {356--362} (\bibinfo {year} {1985})}\BibitemShut {NoStop}%
\bibitem [{\citenamefont {Counsell}, \citenamefont {Levitt},\ and\
  \citenamefont {Ernst}(1985)}]{Counsell1985}%
  \BibitemOpen
  \bibfield  {author} {\bibinfo {author} {\bibfnamefont {C.}~\bibnamefont
  {Counsell}}, \bibinfo {author} {\bibfnamefont {M.}~\bibnamefont {Levitt}}, \
  and\ \bibinfo {author} {\bibfnamefont {R.}~\bibnamefont {Ernst}},\ }\bibfield
   {title} {\enquote {\bibinfo {title} {Analytical theory of composite
  pulses},}\ }\href {\doibase 10.1016/0022-2364(85)90160-x} {\bibfield
  {journal} {\bibinfo  {journal} {Journal of Magnetic Resonance (1969)}\
  }\textbf {\bibinfo {volume} {63}},\ \bibinfo {pages} {133--141} (\bibinfo
  {year} {1985})}\BibitemShut {NoStop}%
\bibitem [{\citenamefont {Tan}\ \emph {et~al.}(2015)\citenamefont {Tan},
  \citenamefont {Rajeswari}, \citenamefont {Madhu},\ and\ \citenamefont
  {Ernst}}]{Tan2015a}%
  \BibitemOpen
  \bibfield  {author} {\bibinfo {author} {\bibfnamefont {K.~O.}\ \bibnamefont
  {Tan}}, \bibinfo {author} {\bibfnamefont {M.}~\bibnamefont {Rajeswari}},
  \bibinfo {author} {\bibfnamefont {P.~K.}\ \bibnamefont {Madhu}}, \ and\
  \bibinfo {author} {\bibfnamefont {M.}~\bibnamefont {Ernst}},\ }\bibfield
  {title} {\enquote {\bibinfo {title} {{Asynchronous symmetry-based sequences
  for homonuclear dipolar recoupling in solid-state nuclear magnetic
  resonance}},}\ }\href {\doibase 10.1063/1.4907275} {\bibfield  {journal}
  {\bibinfo  {journal} {J. Chem. Phys.}\ }\textbf {\bibinfo {volume} {142}},\
  \bibinfo {pages} {1--9} (\bibinfo {year} {2015})}\BibitemShut {NoStop}%
\bibitem [{\citenamefont {Vega}(1978)}]{Vega1978}%
  \BibitemOpen
  \bibfield  {author} {\bibinfo {author} {\bibfnamefont {S.}~\bibnamefont
  {Vega}},\ }\bibfield  {title} {\enquote {\bibinfo {title} {Fictitious spin
  1/2 operator formalism for multiple quantum {NMR}},}\ }\href {\doibase
  10.1063/1.435679} {\bibfield  {journal} {\bibinfo  {journal} {The Journal of
  Chemical Physics}\ }\textbf {\bibinfo {volume} {68}},\ \bibinfo {pages}
  {5518--5527} (\bibinfo {year} {1978})}\BibitemShut {NoStop}%
\bibitem [{\citenamefont {Wokaun}\ and\ \citenamefont
  {Ernst}(1977)}]{Wokaun1977}%
  \BibitemOpen
  \bibfield  {author} {\bibinfo {author} {\bibfnamefont {A.}~\bibnamefont
  {Wokaun}}\ and\ \bibinfo {author} {\bibfnamefont {R.~R.}\ \bibnamefont
  {Ernst}},\ }\bibfield  {title} {\enquote {\bibinfo {title} {Selective
  excitation and detection in multilevel spin systems: Application of single
  transition operators},}\ }\href {\doibase 10.1063/1.435038} {\bibfield
  {journal} {\bibinfo  {journal} {The Journal of Chemical Physics}\ }\textbf
  {\bibinfo {volume} {67}},\ \bibinfo {pages} {1752--1758} (\bibinfo {year}
  {1977})}\BibitemShut {NoStop}%
\bibitem [{\citenamefont {Baum}, \citenamefont {Tycko},\ and\ \citenamefont
  {Pines}(1985)}]{Baum1985}%
  \BibitemOpen
  \bibfield  {author} {\bibinfo {author} {\bibfnamefont {J.}~\bibnamefont
  {Baum}}, \bibinfo {author} {\bibfnamefont {R.}~\bibnamefont {Tycko}}, \ and\
  \bibinfo {author} {\bibfnamefont {A.}~\bibnamefont {Pines}},\ }\bibfield
  {title} {\enquote {\bibinfo {title} {{Broadband and adiabatic inversion of a
  two-level system by phase-modulated pulses}},}\ }\href {\doibase
  10.1103/PhysRevA.32.3435} {\bibfield  {journal} {\bibinfo  {journal} {Phys.
  Rev. A}\ }\textbf {\bibinfo {volume} {32}},\ \bibinfo {pages} {3435--3447}
  (\bibinfo {year} {1985})}\BibitemShut {NoStop}%
\bibitem [{\citenamefont {Jeschke}, \citenamefont {Pribitzer},\ and\
  \citenamefont {Doll}(2015)}]{Jeschke2015}%
  \BibitemOpen
  \bibfield  {author} {\bibinfo {author} {\bibfnamefont {G.}~\bibnamefont
  {Jeschke}}, \bibinfo {author} {\bibfnamefont {S.}~\bibnamefont {Pribitzer}},
  \ and\ \bibinfo {author} {\bibfnamefont {A.}~\bibnamefont {Doll}},\
  }\bibfield  {title} {\enquote {\bibinfo {title} {{Coherence Transfer by
  Passage Pulses in Electron Paramagnetic Resonance Spectroscopy}},}\ }\href
  {\doibase 10.1021/acs.jpcb.5b02964} {\bibfield  {journal} {\bibinfo
  {journal} {J. Phys. Chem. B}\ }\textbf {\bibinfo {volume} {119}},\ \bibinfo
  {pages} {13570--13582} (\bibinfo {year} {2015})}\BibitemShut {NoStop}%
\bibitem [{\citenamefont {Hogben}\ \emph {et~al.}(2011)\citenamefont {Hogben},
  \citenamefont {Krzystyniak}, \citenamefont {Charnock}, \citenamefont {Hore},\
  and\ \citenamefont {Kuprov}}]{Hogben2011}%
  \BibitemOpen
  \bibfield  {author} {\bibinfo {author} {\bibfnamefont {H.}~\bibnamefont
  {Hogben}}, \bibinfo {author} {\bibfnamefont {M.}~\bibnamefont {Krzystyniak}},
  \bibinfo {author} {\bibfnamefont {G.}~\bibnamefont {Charnock}}, \bibinfo
  {author} {\bibfnamefont {P.}~\bibnamefont {Hore}}, \ and\ \bibinfo {author}
  {\bibfnamefont {I.}~\bibnamefont {Kuprov}},\ }\bibfield  {title} {\enquote
  {\bibinfo {title} {Spinach {\textendash} a software library for simulation of
  spin dynamics in large spin systems},}\ }\href {\doibase
  10.1016/j.jmr.2010.11.008} {\bibfield  {journal} {\bibinfo  {journal}
  {Journal of Magnetic Resonance}\ }\textbf {\bibinfo {volume} {208}},\
  \bibinfo {pages} {179--194} (\bibinfo {year} {2011})}\BibitemShut {NoStop}%
\bibitem [{\citenamefont {Levitt}\ and\ \citenamefont
  {Bari}(1992)}]{Levitt1992}%
  \BibitemOpen
  \bibfield  {author} {\bibinfo {author} {\bibfnamefont {M.~H.}\ \bibnamefont
  {Levitt}}\ and\ \bibinfo {author} {\bibfnamefont {L.~D.}\ \bibnamefont
  {Bari}},\ }\bibfield  {title} {\enquote {\bibinfo {title} {Steady state in
  magnetic resonance pulse experiments},}\ }\href {\doibase
  10.1103/physrevlett.69.3124} {\bibfield  {journal} {\bibinfo  {journal}
  {Physical Review Letters}\ }\textbf {\bibinfo {volume} {69}},\ \bibinfo
  {pages} {3124--3127} (\bibinfo {year} {1992})}\BibitemShut {NoStop}%
\bibitem [{\citenamefont {Lebedev}\ and\ \citenamefont
  {Laikov}(1999)}]{Lebedev1999}%
  \BibitemOpen
  \bibfield  {author} {\bibinfo {author} {\bibfnamefont {V.}~\bibnamefont
  {Lebedev}}\ and\ \bibinfo {author} {\bibfnamefont {D.}~\bibnamefont
  {Laikov}},\ }\bibfield  {title} {\enquote {\bibinfo {title} {Quadrature
  formula for the sphere of 131-th algebraic order of accuracy},}\ }\href@noop
  {} {\bibfield  {journal} {\bibinfo  {journal} {Dokl. Akad. Nauk SSSR}\
  }\textbf {\bibinfo {volume} {366}},\ \bibinfo {pages} {741--745} (\bibinfo
  {year} {1999})}\BibitemShut {NoStop}%
\bibitem [{\citenamefont {Doll}\ and\ \citenamefont
  {Jeschke}(2017)}]{Doll2017}%
  \BibitemOpen
  \bibfield  {author} {\bibinfo {author} {\bibfnamefont {A.}~\bibnamefont
  {Doll}}\ and\ \bibinfo {author} {\bibfnamefont {G.}~\bibnamefont {Jeschke}},\
  }\bibfield  {title} {\enquote {\bibinfo {title} {{Wideband frequency-swept
  excitation in pulsed EPR spectroscopy}},}\ }\href {\doibase
  10.1016/j.jmr.2017.01.004} {\bibfield  {journal} {\bibinfo  {journal}
  {Journal of Magnetic Resonance}\ }\textbf {\bibinfo {volume} {280}},\
  \bibinfo {pages} {46--62} (\bibinfo {year} {2017})}\BibitemShut {NoStop}%
\bibitem [{\citenamefont {Tan}\ \emph {et~al.}(2014)\citenamefont {Tan},
  \citenamefont {Nielsen}, \citenamefont {Meier},\ and\ \citenamefont
  {Ernst}}]{Tan2014}%
  \BibitemOpen
  \bibfield  {author} {\bibinfo {author} {\bibfnamefont {K.~O.}\ \bibnamefont
  {Tan}}, \bibinfo {author} {\bibfnamefont {A.~B.}\ \bibnamefont {Nielsen}},
  \bibinfo {author} {\bibfnamefont {B.~H.}\ \bibnamefont {Meier}}, \ and\
  \bibinfo {author} {\bibfnamefont {M.}~\bibnamefont {Ernst}},\ }\bibfield
  {title} {\enquote {\bibinfo {title} {Broad-band {DREAM} recoupling
  sequence},}\ }\href {\doibase 10.1021/jz501703e} {\bibfield  {journal}
  {\bibinfo  {journal} {The Journal of Physical Chemistry Letters}\ }\textbf
  {\bibinfo {volume} {5}},\ \bibinfo {pages} {3366--3372} (\bibinfo {year}
  {2014})}\BibitemShut {NoStop}%
\bibitem [{\citenamefont {Scheuer}\ and\ \citenamefont
  {Naydenov}(2020)}]{Scheuer2020}%
  \BibitemOpen
  \bibfield  {author} {\bibinfo {author} {\bibfnamefont {J.}~\bibnamefont
  {Scheuer}}\ and\ \bibinfo {author} {\bibfnamefont {B.}~\bibnamefont
  {Naydenov}},\ }\bibfield  {title} {\enquote {\bibinfo {title} {Dynamic
  nuclear polarization ({DNP}) in diamond},}\ }in\ \href {\doibase
  10.1016/bs.semsem.2020.03.009} {\emph {\bibinfo {booktitle} {Diamond for
  Quantum Applications Part 1}}}\ (\bibinfo  {publisher} {Elsevier},\ \bibinfo
  {year} {2020})\ pp.\ \bibinfo {pages} {277--293}\BibitemShut {NoStop}%
\bibitem [{\citenamefont {Jeschke}, \citenamefont {Rakhmatullin},\ and\
  \citenamefont {Schweiger}(1998)}]{Jeschke1998a}%
  \BibitemOpen
  \bibfield  {author} {\bibinfo {author} {\bibfnamefont {G.}~\bibnamefont
  {Jeschke}}, \bibinfo {author} {\bibfnamefont {R.}~\bibnamefont
  {Rakhmatullin}}, \ and\ \bibinfo {author} {\bibfnamefont {A.}~\bibnamefont
  {Schweiger}},\ }\bibfield  {title} {\enquote {\bibinfo {title} {{Sensitivity
  Enhancement by Matched Microwave Pulses in One- and Two-Dimensional Electron
  Spin Echo Envelope Modulation Spectroscopy}},}\ }\href {\doibase
  10.1006/jmre.1998.1367} {\bibfield  {journal} {\bibinfo  {journal} {Journal
  of magnetic resonance}\ }\textbf {\bibinfo {volume} {131}},\ \bibinfo {pages}
  {261--271} (\bibinfo {year} {1998})}\BibitemShut {NoStop}%
\bibitem [{\citenamefont {Rizzato}\ \emph {et~al.}(2013)\citenamefont
  {Rizzato}, \citenamefont {Kaminker}, \citenamefont {Vega},\ and\
  \citenamefont {Bennati}}]{Rizzato2013}%
  \BibitemOpen
  \bibfield  {author} {\bibinfo {author} {\bibfnamefont {R.}~\bibnamefont
  {Rizzato}}, \bibinfo {author} {\bibfnamefont {I.}~\bibnamefont {Kaminker}},
  \bibinfo {author} {\bibfnamefont {S.}~\bibnamefont {Vega}}, \ and\ \bibinfo
  {author} {\bibfnamefont {M.}~\bibnamefont {Bennati}},\ }\bibfield  {title}
  {\enquote {\bibinfo {title} {{Cross-polarisation edited ENDOR}},}\ }\href
  {\doibase 10.1080/00268976.2013.816795} {\bibfield  {journal} {\bibinfo
  {journal} {Mol. Phys.}\ }\textbf {\bibinfo {volume} {111}},\ \bibinfo {pages}
  {2809--2823} (\bibinfo {year} {2013})}\BibitemShut {NoStop}%
\end{thebibliography}%

\end{document}


\lstset{language=Matlab,%
	basicstyle=\ttfamily,
	breaklines=true,%
	morekeywords={matlab2tikz},
	keywordstyle=\color{blue},%
	morekeywords=[2]{1}, keywordstyle=[2]{\color{black}},
	identifierstyle=\color{black},%
	stringstyle=\color{mylilas},
	commentstyle=\color{mygreen},%
	showstringspaces=false,
	numbers=left,%
	numberstyle={\tiny \color{black}},
	numbersep=9pt, 
	emph=[1]{for,end,break},emphstyle=[1]\color{red}, 
}

\pagestyle{fancy}
\thispagestyle{plain}
\fancypagestyle{plain}



\makeFNbottom
\makeatletter
\renewcommand\LARGE{\@setfontsize\LARGE{15pt}{17}}
\renewcommand\Large{\@setfontsize\Large{12pt}{14}}
\renewcommand\large{\@setfontsize\large{10pt}{12}}
\renewcommand\footnotesize{\@setfontsize\footnotesize{7pt}{10}}
\makeatother

\renewcommand{\thefootnote}{\fnsymbol{footnote}}
\renewcommand\footnoterule{\vspace*{1pt}%
\color{cream}\hrule width 3.5in height 0.4pt \color{black}\vspace*{5pt}} 
\setcounter{secnumdepth}{5}

\makeatletter 
\renewcommand\@biblabel[1]{#1}            
\renewcommand\@makefntext[1]%
{\noindent\makebox[0pt][r]{\@thefnmark\,}#1}
\makeatother 
\renewcommand{\figurename}{\small{Fig.}~}
\sectionfont{\sffamily\Large}
\subsectionfont{\normalsize}
\subsubsectionfont{\bf}
\setstretch{1.125} 
\setlength{\skip\footins}{0.8cm}
\setlength{\footnotesep}{0.25cm}
\setlength{\jot}{10pt}
\titlespacing*{\section}{0pt}{4pt}{4pt}
\titlespacing*{\subsection}{0pt}{15pt}{1pt}

\fancyfoot{}
\fancyfoot[LO,RE]{}
\fancyfoot[CO]{}
\fancyfoot[CE]{}
\fancyfoot[R]{\footnotesize{SI page \hspace{2pt}\thepage~/~\pageref{lastpage}}}
\fancyhead{}
\renewcommand{\headrulewidth}{0pt} 
\renewcommand{\footrulewidth}{0pt}
\setlength{\arrayrulewidth}{1pt}
\setlength{\columnsep}{6.5mm}
\setlength\bibsep{1pt}

\makeatletter 
\newlength{\figrulesep} 
\setlength{\figrulesep}{0.5\textfloatsep} 

\newcommand{\topfigrule}{\vspace*{-1pt}%
\noindent{\color{cream}\rule[-\figrulesep]{\columnwidth}{1.5pt}} }

\newcommand{\botfigrule}{\vspace*{-2pt}%
\noindent{\color{cream}\rule[\figrulesep]{\columnwidth}{1.5pt}} }

\newcommand{\dblfigrule}{\vspace*{-1pt}%
\noindent{\color{cream}\rule[-\figrulesep]{\textwidth}{1.5pt}} }

\makeatother



\renewcommand*\rmdefault{bch}\normalfont\upshape
\rmfamily

\onecolumn

\title{Supporting Information for:\\
	Designing Broadband Pulsed Dynamic Nuclear Polarization Sequences in Static Solids}
\date{}
\author{}
\maketitle

  {\centering
  	\large{Nino Wili\textit{$^{1}$}, Anders Bodholt Nielsen\textit{$^{2}$}, Laura Alicia Völker\textit{$^{1}$}, Lukas Schreder\textit{$^{1}$}, Niels Chr. Nielsen\textit{$^{2}$}, Gunnar Jeschke\textit{$^{1}$} and Kong Ooi Tan\textit{$^{3}$}} \\

\vspace*{1cm}

\textit{$^{1}$~Department of Chemistry and Applied Biosciences, Laboratory of Physical Chemistry, ETH Zurich,
		Vladimir-Prelog-Weg 2, 8093 Zurich, Switzerland. E-mail: nino.wili@alumni.ethz.ch}\\
\vspace*{5mm}
	
{\textit{$^{2}$~Interdisciplinary Nanoscience Center (iNANO) and Department of Chemistry, Aarhus University, Gustav Wieds Vej 14, DK-8000 Aarhus C, Denmark}}
\\
\vspace*{5mm}

{\textit{$^{3}$~Laboratoire des Biomolécules, LBM,  Département de Chimie, École Normale Supérieure, PSL University, Sorbonne Université, CNRS, 75005 Paris, France. E-mail: kong-ooi.tan@ens.psl.eu}}

}

\vspace*{1cm}
\setcounter{tocdepth}{1} 
\tableofcontents

\clearpage


\section{Calculation of Fourier coefficients and scaling factors for BASE}  

In this section, we show an example of how to calculate the effective fields and scaling factors on the example of BASE. Although BASE is best applied on-resonant, we include here an offset of 5~MHz for illustration. The example can be found in the file \texttt{DNPexample\_BASE.m}.

\subsection{Build the rf-irradiation}

\begin{lstlisting}[firstline=1,lastline=35]
clear, close all

% add helper functions
addpath(genpath('./core/'))


%% define sequence parameters
nu_I=-14.83;            % Nuclear Zeeman frequency
nu_1=32;                  % Electron Rabi freq
tp1=20*1e-3;            % first pulse length in BASE
tp2=(1+tp1*(nu_1-abs(nu_I)))/(abs(nu_I)+nu_1);             % second pulse length in BASE


nu1_vec = nu_1*[1 1];
tp_vec = [tp1 tp2];
phi_vec = [0 180]/180*pi; % phases of the pulses
dt =0.1e-4;             % time step of numerical IFT

offset = 5;

rho0_vec = [-1 0 0]';

%% build rf
[ rf, time ]=build_rf(tp_vec,nu1_vec,phi_vec,dt);

%plotting
h = figure(1);
clf
hold on
plot(time*1e3,real(rf),'b')
plot(time*1e3,imag(rf),'r')
xlabel('t / ns')
ylabel('\nu_1 / MHz')
axis([-1 60 -40 40])
legend('real(rf)','imag(rf)','location','northeast')
\end{lstlisting}

\begin{figure}[hbt!]
	\centering
	\includegraphics[width=12cm]{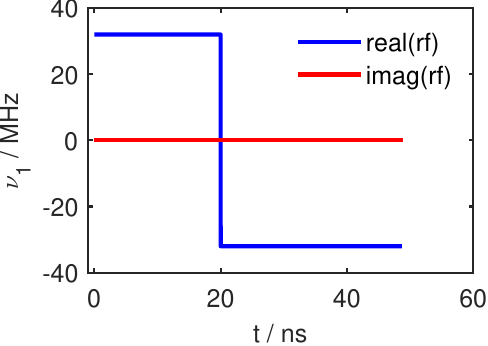}
	\caption{rf-irradiation scheme for the BASE simulation.} 
\end{figure}

\clearpage

\subsection{Calculate $R^\text{(control)}(t)$}

\begin{lstlisting}[firstnumber=42,firstline=42,lastline=75]
% generate a vector of offsets the same length as rf
offset=offset*ones(size(rf));

%calculate modulation frequency
wmod=2*pi/(time(end)+dt);
nu_m = wmod/2/pi;


% pre-allocate rotation matrices
R_control = zeros(3,3,numel(time));
R_flipped = zeros(3,3,numel(time));


% build array of rotation quaternions for every step
q_pulse = zeros(4,numel(time));
for it = 1:numel(time)
    q_pulse(:,it)=quat_rf(abs(rf(it)),angle(rf(it)),offset(it),dt,'frame') ;
end


% multiply the quaternionas step-by-step and
% express them as an array of rotation matrices (for plotting only at this stage)
qtot = [1 0 0 0]';
q_control = zeros(4,numel(time));
for it=1:numel(time)
    q_control(:,it)=qtot;
    R_control(:,:,it) = quat2rotmat(qtot);
    qtot = quatmult(qtot,q_pulse(:,it));
end



% Plot IF in rotating frame
IF_plot(time,R_control,2);
\end{lstlisting}
\begin{figure}[hbt!]
	\centering
	\includegraphics[width=12cm]{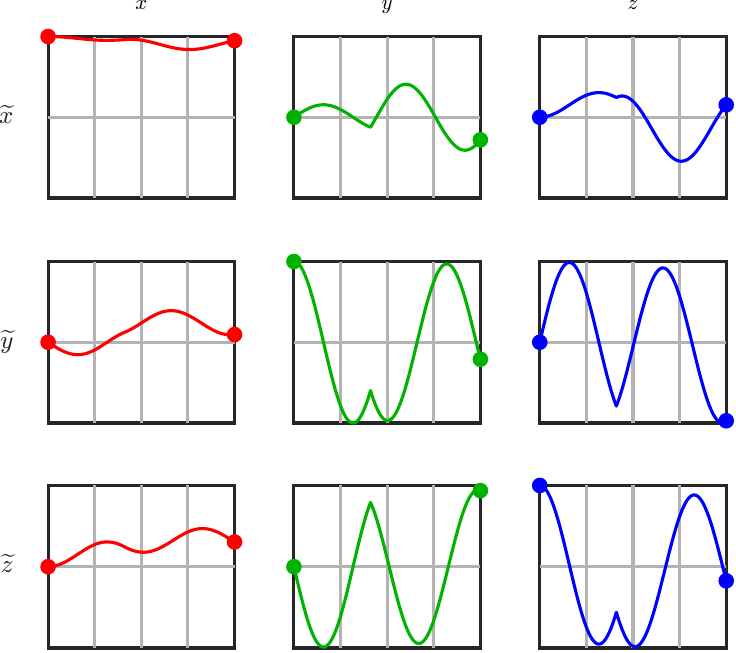}
	\caption{Time-dependence of each element of $R^\text{(control)}(t)$.} 
\end{figure}

\clearpage

\subsection{Remove the effective field from $R^\text{(control)}(t)$, determine $R^\text{(eff)}(t)$}

\begin{lstlisting}[firstnumber=82,firstline=82,lastline=123]
%% flip each quaternion of q_control (or R_control) 
% such that the effective field is now along the z-axis

%determine the effective field (effective flip angle and direction)
[z_eff,beta_eff] = quat2eff(qtot);

% determine R_flip, or the corresponding quaternion, respectively
% see https://doi.org/10.1063/1.5123046 for details
u_z = [0 0 1]';
q_flip = cross(z_eff,u_z);
qr_flip = 1+u_z'*z_eff;
q_flip = [qr_flip; q_flip ];
q_flip = q_flip/sqrt(sum(q_flip'*q_flip));

% flip all quaternions/rot-matrices
q_flipped= zeros(4,numel(time));
for it=1:numel(time)
    q_flipped(:,it)=quatmult(q_flip,q_control(:,it));
    R_flipped(:,:,it) = quat2rotmat(q_flipped(:,it));
end

%% remove the effective field from the IFT, i.e. calculate R_eff(t)
% determine the effective field frequency
w_eff = beta_eff/(time(end)+dt);
nu_eff = w_eff/(2*pi);

%pre-allocate rotation quaternions/matrices
q_eff = zeros(4,numel(time));
R_eff = zeros(3,3,numel(time));

% calculate R_eff(t)
for it=1:numel(time)
    
    %calculate Rz(-weff*t), 
    beta = -w_eff*time(it);
    q_z = [cos(beta/2) sin(beta/2)*[0 0 1]]';
    
    q_eff(:,it)=quatmult(q_z,q_flipped(:,it));
    R_eff(:,:,it) = quat2rotmat(q_eff(:,it));
end

IF_plot(time,R_eff,3);
\end{lstlisting}
\begin{figure}[hbt!]
	\centering
	\includegraphics[width=12cm]{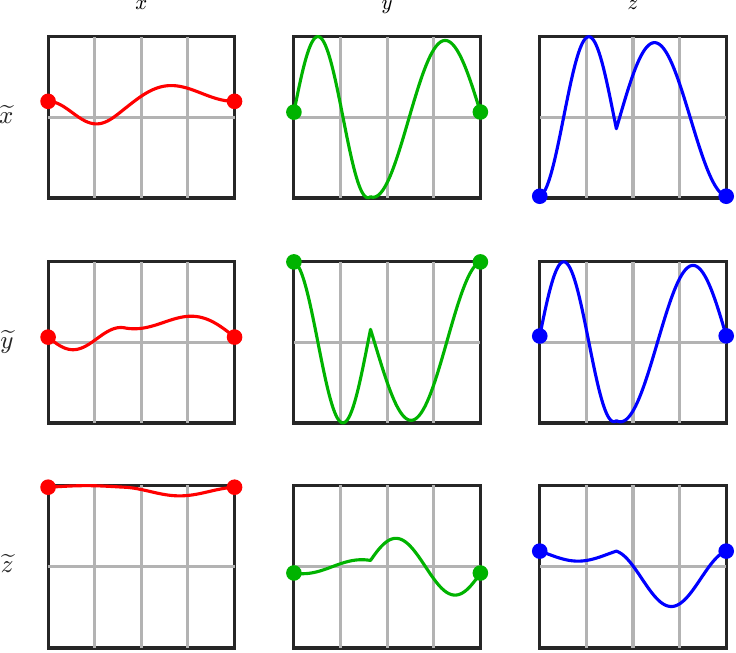}
	\caption{Time-dependence of each element of $R^\text{(eff)}(t)$. Note that all elements are preiodic.} 
\end{figure}

\clearpage

\subsection{Calculate the Fourier coefficients and scaling factors}
\begin{lstlisting}[firstnumber=128,firstline=128,lastline=181]
%%  calculate Fourier coefficients a^(k)
N=numel(time);
k_vec = ((0:N-1)-fix(N/2));
if mod(N,2)==0
    k_vec = k_vec(2:end);
end

A_k = zeros(3,3,numel(k_vec));


% Fourier transform each element (xx,xy,xz,...zz) of R_eff(t)
for ii=1:3
    for jj=1:3
        y = squeeze(R_eff(ii,jj,:));
        a = fftshift(fft(y))/numel(y);
        if mod(N,2)==0
            A_k(ii,jj,:)=a(2:end);
        else
            A_k(ii,jj,:)=a(1:end);
        end
    end
end


%extract the original z operator (should no oszillate during free evolution)
Az_k = squeeze(A_k(:,3,:));

%% extract the relevant scaling factor

[nu_eff_I,kI] = get_nu_eff_I(nu_I,nu_m);

kmax=max(k_vec);

ax = squeeze(Az_k(1,:));
ay = squeeze(Az_k(2,:));
ap = ax-1i*ay;
am = ax+1i*ay;
az = squeeze(Az_k(3,:));


k0 = kmax+1;
k = kmax+1+kI;
mk = kmax+1-kI;

rel_sign = sign(nu_eff*nu_eff_I);
if rel_sign == 1
    a_eff= -sqrt(ap(mk)*am(k)) ;
else
    a_eff= sqrt(am(mk)*ap(k)) ;
end

proj = z_eff'*rho0_vec;

f_pm = a_eff*proj
\end{lstlisting}

\subsection{Compare with fully numerical calculation}
\begin{lstlisting}[firstnumber=183,firstline=183,lastline=230]
%% comparison with numerical simulation

Ntheta = 31; %number of orientations

r=4.5; % e-n distance in Angstrom
Nrounds = 200; %number of repetitions of the DNP element

%caluclate the transfer numerically
[t_num,sig_num]=DNP_numerical(tp_vec,phi_vec,nu1_vec,offset,nu_I,rho0_vec,r,Ntheta,Nrounds);

h4=figure(4);
clf
hold on
plot(t_num,sig_num)

%% analytical calculation
%calculate the mismatch
delta_nu_eff = abs(nu_eff_piecewise(tp_vec,phi_vec,nu1_vec,offset))-abs(nu_eff_I);

%calculate the anisotropy of the hf-coupling
natural_constants
T = mu0/(4*pi)*gfree*bmagn*g1H*nmagn*1/(r*1e-10)^3/planck/1e6;

%generate orientations and corresponding weights
theta_vec = linspace(0,pi/2,Ntheta);
weights=sin(theta_vec); weights=weights/sum(weights);

%preallocate results
t_theo = t_num(1:10:end);
sig_theo = zeros(1,numel(t_theo));

%loop over orientations, sum up results
for itheta = 1:Ntheta
    
    theta=theta_vec(itheta);
    B=3*sin(theta)*cos(theta)*T;
    nu_pm = sqrt(B^2*a_eff^2/4+delta_nu_eff^2); %transfer frequency
    transfer_amp = a_eff^2*B^2/(4*nu_pm^2); %amplitude of transfer
    
    sig_theo = sig_theo+weights(itheta)*sign(a_eff)*proj*transfer_amp*...
                sin(1/2*(2*pi*nu_pm)*t_theo).^2;
end
plot(t_theo,sig_theo,'o')

axis([xlim -0.1 0.8])
xlabel('t / \mus')
ylabel('<I_z>')
legend('num.','theo.','location','northeast')
\end{lstlisting}
\begin{figure}[hbt!]
	\centering
	\includegraphics[width=12cm]{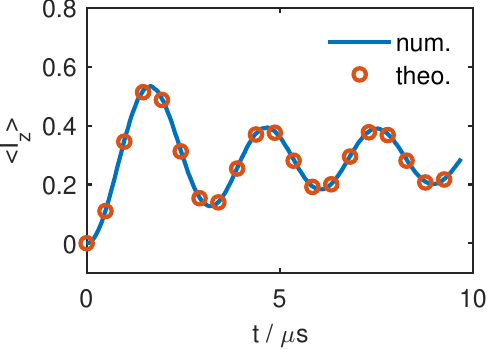}
	\caption{Fully numerical simulation vs. effective Hamiltonian calculation.} 
\end{figure}

\clearpage
\section{Comparison of numerical and analytical computations for strongly coupled protons}

\begin{figure}[hbt!]
	\centering
	\includegraphics[width=12cm]{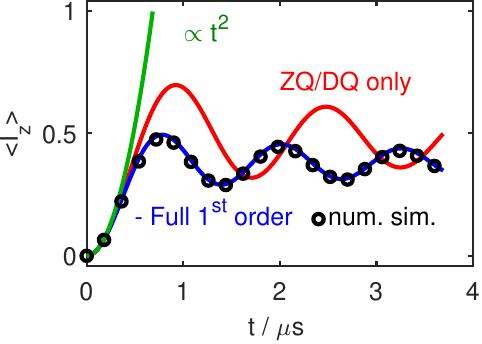}
	\caption{Comparison of $S_z \rightarrow I_z$ polarization transfer efficiencies calculated using an effective Hamiltonian including all first order terms (blue) or only the flip-flop terms (red) 
		with a full numerical simulation (black circles). A two-spin e-$^1$H spin pair with a distance $r_{en}$ = 2.5 Å is used in the numerical simulations. This is a very strongly coupled proton.	The green line illustrates the initial build-up} 
		\label{fig:close proton}
\end{figure}

Figure 4(b) in the main text showed a comparison of numerical and analytical computations for a proton at a distance of $r_{en}$ = 4.5 Å. This is is already quite close. Here we show what happens if one looks at protons that are even closer (2.5 Å ), see Figure \ref{fig:close proton}. In this case, the hyperfine coupling is very strong. Nevertheless, the full first-order Hamiltonian (blue line) still perfectly describes the transfer. However, it is clear that only looking at the effective fields and the flip-flop terms (red) becomes more and more problematic for strong hyperfine couplings.

\clearpage

\section{Three-spin transitions}
\begin{figure}[hbt!]
	\centering
	\includegraphics[width=12cm]{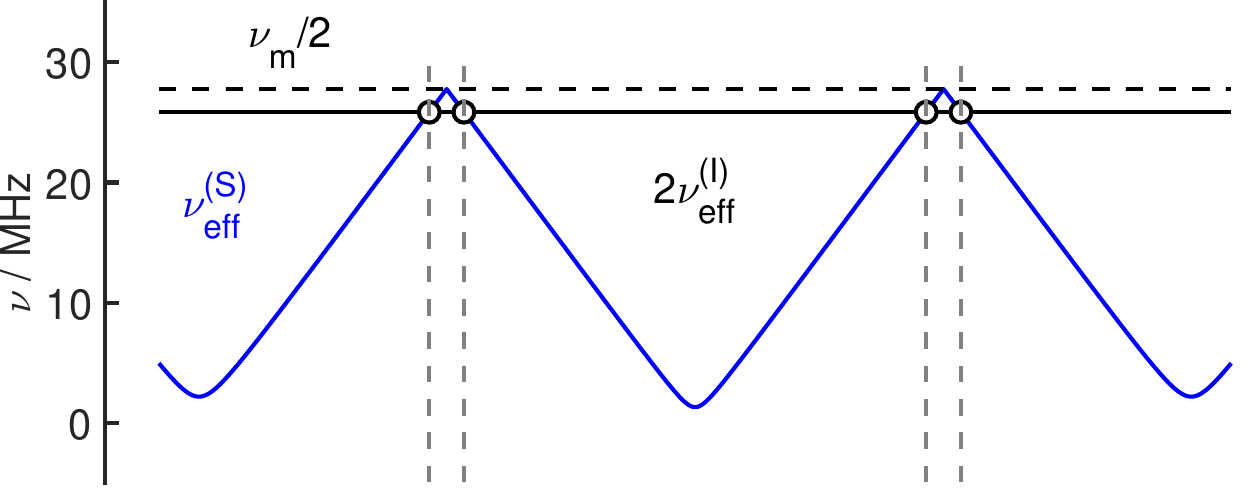}
	\includegraphics[width=12cm]{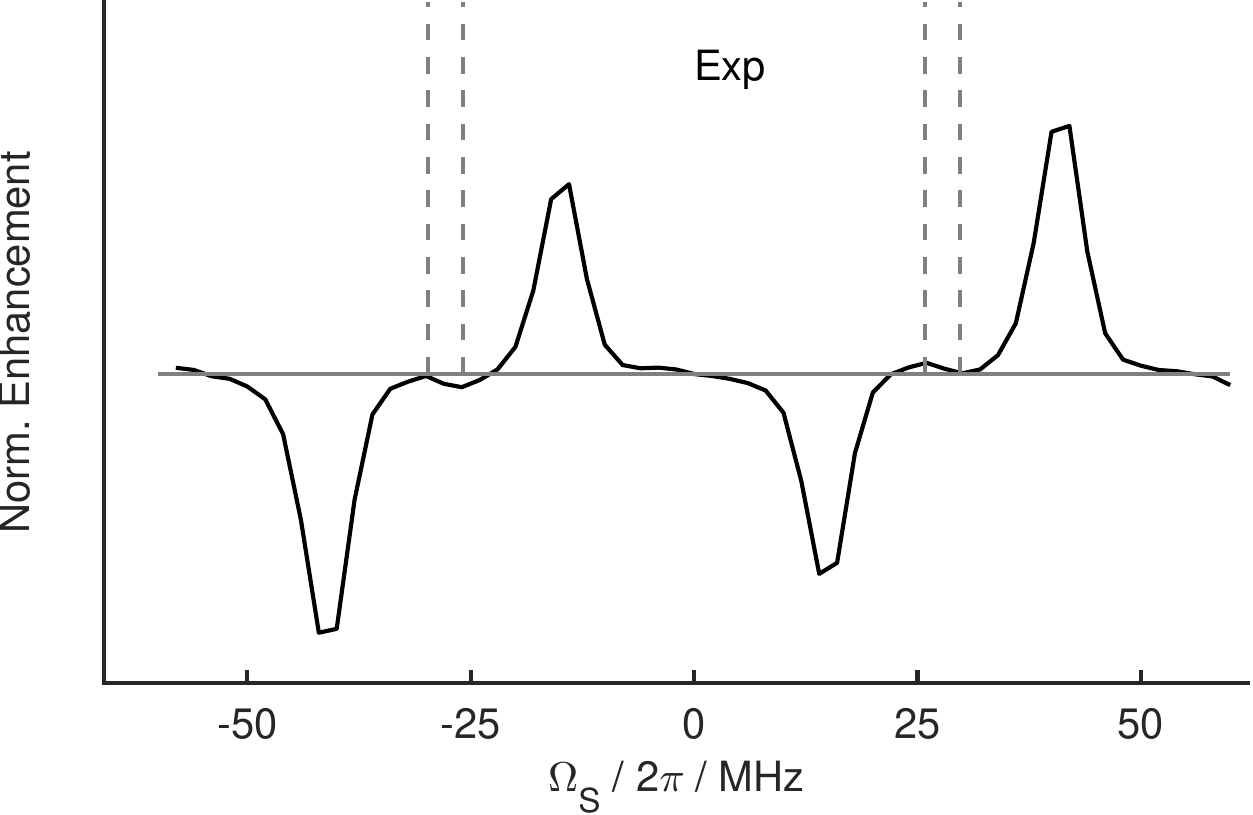}
	\caption{Predicted offsets for three-spin transitions for XiX DNP with $t_{p,1}$=12~ns and $t_{p,2}$=6~ns.} 
	\label{fig:three_spin}
\end{figure}

Some small features in the Experimental DNP profiles cannot be explained by the electron-nuclear two-spin model. It is well known that there are also electron-nuclear-nuclear three-spin transitions. A comprehensive treatment for them would require second-order average Hamiltonian theory. However, the position of these features can be estimated by simply doubling the nuclear Zeeman frequency and then looking for matching conditions in exactly the same way as in the two-spin case. This is shown in Fig. \ref{fig:three_spin}.

\clearpage

\section{Resonator profile}

\begin{figure}[hbt!]
	\centering
	\includegraphics[width=12cm]{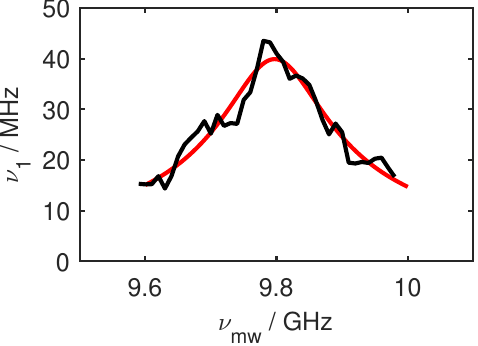}
	\caption{Experimental resonator profile (black), determined by nutation experiments, and Lorentzian fit (red). $\nu_{1,\text{max}}$=40~MHz, $Q$=61.} 
	\label{fig:resonator}
\end{figure}
\clearpage

\section{Experimental data for the adiabatic solid effect (ASE)}
\begin{figure}[hbt!]
	\centering
	\includegraphics[width=12cm]{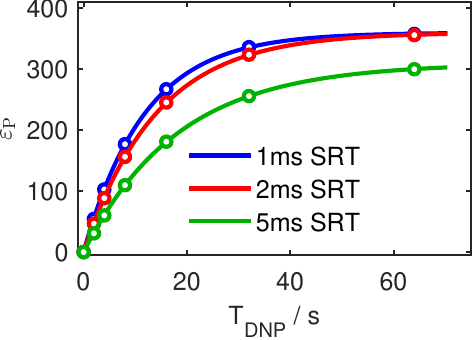}
	\caption{Polarization build-up using the adiabatic solid effect with different repetition times.} 
	\label{fig:ASE}
\end{figure}

\begin{table}[hbt!]
	\ra{1.3}
	\caption{Enhancements $\varepsilon_\mathrm{max}$, build-up times $T_\mathrm{B}$ , and sensitivity per unit time (i.e. signal per square root of time) $\varepsilon_\mathrm{max}\cdot\sqrt{{T_\mathrm{1,n}}/{T_\mathrm{B}}}$ for ASE. $T_\mathrm{1,n}=$~36.2~s. $T_\mathrm{1,e}=$~2.5~ms.}
	\centering
		\begin{tabular}{l c c c }
			\toprule
			& \multicolumn{3}{c}{ ASE} \\
			\midrule 
			$t_\text{rep}$ & 1~ms & 2~ms & 5~ms\\
			\midrule
				$\varepsilon_\mathrm{max}$ & 360 & 360  & 309 \\
				$T_\mathrm{B}$ / s& 11.9  & 14.1 & 18.3 \\
				$\varepsilon_\mathrm{max}\cdot\sqrt{\frac{T_\mathrm{1,n}}{T_\mathrm{B}}}$& 629 & 578 & 435 \\
			
			\bottomrule
		\end{tabular}
	\label{tab:enhancements}
\end{table}

\clearpage
\section{Experimental determination of $T_1,e$}
\begin{figure}[hbt!]
	\centering
	\includegraphics[width=12cm]{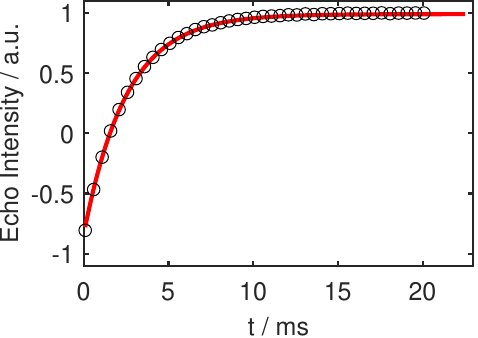}
	\caption{Inversion recovery data for the electron spin (black circles) and exponential fit (red). $T_{1,e}$=2.5~ms}. 
	\label{fig:T1e}
\end{figure}
\clearpage

\section{Experimental determination of $T_1,n$}
\begin{figure}[hbt!]
	\centering
	\includegraphics[width=12cm]{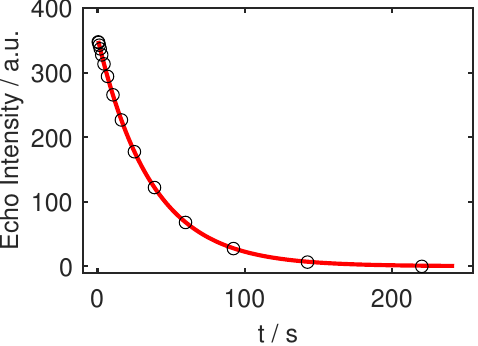}
	\caption{Experimental proton polarization decay after 60~s of adiabatic solid effect DNP (black circles) and exponential fit (red). $T_{1,n}$=36.2~s}. 
	\label{fig:T1n_DNP}
\end{figure}
\begin{figure}[hbt!]
	\centering
	\includegraphics[width=12cm]{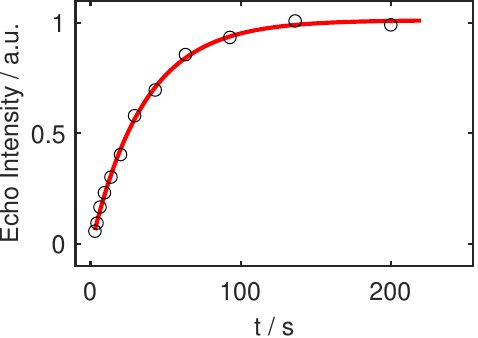}
	\caption{Proton saturation recovery data without DNP (black circles) and exponential fit (red). $T_{1,n}$=35~s}. 
	\label{fig:T1n_satrec}
\end{figure}

\label{lastpage}